\expandafter\edef\csname hypers\string @fe\endcsname{\catcode
                                             `\noexpand @=\the\catcode`\@}%
\catcode`\@=11
%
%
\ifx\hyper@utoprocess\hyper@ndefined
\else
 \expandafter\hyper@utoprocess\fi
\ifx\hyperd@ne\hyper@ndefined
 \global\let\hyperd@ne=\relax
\else
 \hypers@fe
 \errhelp{hyperbasics.tex needs to be included only once outside
          of any {...} or \begingroup...\endgroup. You have tried to
          include it more than once. If the previous include was indeed
          outside any groupings, continue and all will be well.}%
 \errmessage{Input this file only once!}%
 \expandafter \fi
%
%
\def\hyperv@rsion{12}%
%
%
\newread\hyperf@le
\def\hyperf@lename{\jobname.hrf}%
\immediate\openin\hyperf@le\hyperf@lename\relax
\ifeof\hyperf@le\relax
 \immediate\closein\hyperf@le\relax
\else
 \immediate\closein\hyperf@le\relax
 \input \hyperf@lename
\fi
%
%
\newwrite\hyperf@le
\immediate\openout\hyperf@le\hyperf@lename
%
%
\newtoks\hypert@ks
%
%
\edef\hypert@mp{\catcode`\noexpand\%=\the\catcode`\%}
\catcode`\%=12
\def\hyperp@rcent{
\hypert@mp
\edef\hypert@mp{\catcode`\noexpand\#=\the\catcode`\#}%
\catcode`\#=12
\def\hyperh@sh{#}%
\hypert@mp
\let\hypert@mp=\relax
\let\hyper@nd=\relax
\def\hyperbl@nk{ }
\def\hyperstr@pquote#1"#2\hyper@nd{
   #1
   \ifx\hyper@nd#2\hyper@nd
   \else\hyperp@rcent22\hyperstr@pquote#2\hyper@nd\fi}%
\def\hyperstr@pblank#1 #2\hyper@nd{
   #1
   \ifx\hyper@nd#2\hyper@nd
   \else\hyperp@rcent20\hyperstr@pblank#2\hyper@nd\fi}
\long\def\hyper@nchor#1#2{\edef\hyperm@cro{html:<A #1>}%
                          \special\expandafter{\hyperm@cro}%
                          {#2}}%
\def\hyper@atm@ning#1->#2\hyper@nd{#2}
\def\hyperlink{\protect\hyperlink@}
\def\hyperlink@{{\catcode\expandafter`\noexpand\#=12 
                 \catcode\expandafter`\noexpand\%=12 
                 \catcode\expandafter`\noexpand\~=12
                \expandafter}\hyperlink@@}
\def\hyperlink@@#1{\protect\hyperlink@@@{#1}}
\def\hyperlink@@@#1{\edef\hypert@mp{#1}%
               \edef\hypert@mp{\expandafter\hyper@atm@ning\meaning\hypert@mp
                               \hyper@nd}%
               \edef\hypert@mp{\expandafter\hyperstr@pquote%
                               \hypert@mp"\hyper@nd}%
               \edef\hypert@mp{\expandafter\expandafter\expandafter
                               \hyperstr@pblank\expandafter%
                               \hypert@mp\hyperbl@nk\hyper@nd}%
               \hyper@nchor{href=\expandafter"\hypert@mp"}}%
\def\hypertarget#1{\edef\hypert@mp{#1}%
               \edef\hypert@mp{\expandafter\hyper@atm@ning\meaning\hypert@mp
                               \hyper@nd}%
               \edef\hypert@mp{\expandafter\hyperstr@pquote%
                               \hypert@mp"\hyper@nd}%
               \edef\hypert@mp{\expandafter\expandafter\expandafter
                               \hyperstr@pblank\expandafter%
                               \hypert@mp\hyperbl@nk\hyper@nd}%
               \hyper@nchor{name=\expandafter"\hypert@mp"}}%
\def\hyperref{\afterassignment\hyperr@f\let\hyperp@ram}
\def\hyperr@f{\ifx\hyperp@ram{\iffalse}\fi
               \expandafter\expandafter\expandafter\hyperr@@
               \expandafter{%
              \else
               \iffalse}\fi
               \ifx\hyperp@ram\hyper@ndefined
                 \message{Undefined reference}%
                 \def\hyperp@r@m{{}{undefined}{}}%
               \else
                 \edef\hyperp@r@m{\hyperp@ram}%
               \fi
               \expandafter\expandafter\expandafter\hyperr@@
               \expandafter\hyperp@r@m
              \fi}%
\def\hyperr@@#1#2#3{\ifx\hyper@ndefined#1\hyper@ndefined
                    \hypert@ks\expandafter{\hyperh@sh#2.#3}%
                    \else
                     \ifx\hyper@ndefined#2#3\hyper@ndefined
                      \hypert@ks{#1}%
                     \else
                      \def\hypert@mp{#1}%
                      \hypert@ks\expandafter\expandafter\expandafter
                      {\expandafter\hypert@mp\hyperh@sh#2.#3}%
                     \fi
                    \fi
                    \expandafter\hyperlink\expandafter{\the\hypert@ks}}%
\def\hyperdef#1#2#3{{\escapechar=`\\\relax
                     \edef\hyper@t@mp@{\hyperstr@pquote#2.#3"\hyper@nd}%
                     \expandafter\ifx\csname hyperd@\meaning\hyper@t@mp@
                     \endcsname
                     \relax
                     \expandafter\gdef\csname hyperd@\meaning\hyper@t@mp@
                     \endcsname{}%
                     \gdef#1{{}{\hyperstr@pquote#2"\hyper@nd}%
                               {\hyperstr@pquote#3"\hyper@nd}}%
                     \immediate\write\hyperf@le{\def\noexpand#1{{}%
                        {\noexpand\hyperstr@pquote#2"\noexpand\hyper@nd}%
                        {\noexpand\hyperstr@pquote#3"\noexpand\hyper@nd}}}%
                     \xdef\hyper@t@mp@{\global\let\noexpand\hyper@t@mp@=\relax
                                       \noexpand\hypertarget{\hyper@t@mp@}}%
                     \global\hypert@ks={\hyper@t@mp@}%
                     \else
                     \message\expandafter{\expandafter'\hyper@t@mp@' duplicate}%
		     \def\hyper@@tmp@{\hyperdef{#1}{#2}}%
		     \edef\hyper@@tmp@@{{#3@}}%
                     \global\let\hyper@t@mp@=\relax
                     \global\hypert@ks=\expandafter\expandafter
		     \expandafter{\expandafter\hyper@@tmp@\hyper@@tmp@@}%
                     \fi}\the\hypert@ks}%

\def\hyper@nique#1#2#3#4{{\escapechar=`\\\relax
                     \edef\hyper@t@mp@{\hyperstr@pquote#2.#3"\hyper@nd}%
                     \expandafter\ifx\csname hyperd@\meaning\hyper@t@mp@
                     \endcsname
                     \relax
                     \gdef#1{{}{\hyperstr@pquote#2"\hyper@nd}%
                               {\hyperstr@pquote#3"\hyper@nd}}%
                     \global\let\hyper@t@mp@=\relax
                     #4%
                     \else
		     \def\hyper@@tmp@{\hyper@nique{#1}{#2}}%
		     \edef\hyper@@tmp@@{{#3@}}%
                     \global\let\hyper@t@mp@=\relax
   		     \expandafter\hyper@@tmp@\hyper@@tmp@@{#4}%
                     \fi}}%

\ifx\protect\hyper@ndefined\let\protect=\relax\fi
\let\hyper@@@@=\relax
\def\hyper@@{\let\hyper@@@=\relax}%
\hyper@@
\def\hyper@{\relax\let\hyper@@@\noexpand\hyper@\noexpand}%
\def\hyperpr@ref{\hyper@@\hyperref}
\def\hyperpr@link{\hyper@@\hyperlink}
\def\hyperpr@def{\hyper@@\hyperdef}
\let\hyper@marker=\relax
\def\hyper@@tokenize#1\hyper@marker{#1}
\def\hyper@tokenize{\expandafter\endgroup\hyper@@tokenize}
\def\hyperpr@tokenize{\hyper@@\hyper@tokenize}

\edef\href{\begingroup\catcode`\string @=11
            \hyper@\hyperpr@tokenize\hyper@\hyperpr@link
            \hyper@marker}
\let\hname\hypertarget
\def\allowoncemore{\def\hyper@utoprocess{\let\hyper@utoprocess=\hyper@ndefined
                                         \hypers@fe }}%
%
%
\hypers@fe
 

\input amssym.tex 

\def\unredoffs{}
\tolerance=1000\hfuzz=2pt\def\fontflag{cm}
\catcode`\@=11 
\ifx\hyperdef\UNd@FiNeD\def\hyperdef#1#2#3#4{#4}\def\hyperref#1#2#3#4{#4}\def\href#1#2{#2}\fi
\magnification=1200\unredoffs\baselineskip=16pt plus 2pt minus 1pt
\def\Title#1#2{\nopagenumbers\rightline{#1}%
\vskip 1in\centerline{#2}\vskip .5in\pageno=0}
\def\Date#1{\vfill\leftline{#1}\tenpoint\supereject%
\footline={\hss\tenrm\hyperdef\hypernoname{page}\folio\folio\hss}}%

{\count255=\time\divide\count255 by 60 \xdef\hourmin{\number\count255}
 \multiply\count255 by-60\advance\count255 by\time
 \xdef\hourmin{\hourmin:\ifnum\count255<10 0\fi\the\count255}
}
\def\date{\number\day.\number\month.\number\year\ at \hourmin}

\def\draft{\message{ DRAFTMODE }\def\draftdate{{\rm Preliminary draft: \date}}%
\headline={\hfil\draftdate}\writelabels\baselineskip=20pt plus 2pt minus 2pt}
\def\nolabels{\def\wrlabeL##1{}\def\eqlabeL##1{}\def\reflabeL##1{}}
\def\writelabels{\def\wrlabeL##1{\leavevmode\vadjust{\rlap{\smash%
{\line{{\escapechar=` \hfill\rlap{\sevenrm\hskip.03in\string##1}}}}}}}%
\def\eqlabeL##1{{\escapechar-1\rlap{\sevenrm\hskip.05in\string##1}}}%
\def\reflabeL##1{\noexpand\llap{\noexpand\sevenrm\string\string\string##1}}}
\nolabels

\global\newcount\secno \global\secno=0
\global\newcount\meqno \global\meqno=1
\def\s@csym{}

\def\newsec#1\par{\global\advance\secno by1%
{\toks0{#1}\message{(\the\secno. \the\toks0)}}%
\global\subsecno=0\eqnres@t\let\s@csym\secsym\xdef\secn@m{\the\secno}\noindent
{\bf\hyperdef\hypernoname{section}{\the\secno}{\the\secno.} #1}%
\writetoca{{\string\hyperref{}{section}{\the\secno}{\bf \the\secno\quad}} {\bf #1}}\par%
\nobreak\medskip\nobreak\noindent\ignorespaces}
\def\eqnres@t{\xdef\secsym{\the\secno.}\global\meqno=1\bigbreak\bigskip}
\def\sequentialequations{\def\eqnres@t{\bigbreak}}\xdef\secsym{}

\global\newcount\subsecno \global\subsecno=0
\def\subsec#1\par{\global\advance\subsecno by1%
{\toks0{#1}\message{(\s@csym\the\subsecno. \the\toks0)}}%
\global\subsubsecno=0%
\ifnum\lastpenalty>9000\else\bigbreak\fi
\noindent{\it\hyperdef\hypernoname{subsection}{\secn@m.\the\subsecno}%
{\secn@m.\the\subsecno.} #1}\writetoca{\string\hskip1.45cm
{\string\hyperref{}{subsection}{\secn@m.\the\subsecno}{\secn@m.\the\subsecno.}}
{#1}}\par\nobreak\medskip\nobreak\noindent\ignorespaces}

\global\newcount\subsubsecno \global\subsubsecno=0
\def\subsubsec#1\par{\global\advance\subsubsecno by1%
{\toks0{#1}\message{(\secn@m.\the\subsecno.\the\subsubsecno. \the\toks0)}}%
\global\subsubsubsecno=0%
\ifnum\lastpenalty>9000\else\bigbreak\fi
\noindent{\it\hyperdef\hypernoname{subsubsection}{\secn@m.\the\subsecno\the\subsubsecno}%
{\secn@m.\the\subsecno.\the\subsubsecno.} #1}
\par\nobreak\medskip\nobreak\noindent\ignorespaces}

\global\newcount\subsubsubsecno \global\subsubsubsecno=0
\def\subsubsubsec#1\par{\global\advance\subsubsubsecno by1%
{\toks0{#1}\message{(\secn@m.\the\subsecno.\the\subsubsecno.\the\subsubsubsecno \the\toks0)}}%
\ifnum\lastpenalty>9000\else\bigbreak\fi
\noindent{\it\hyperdef\hypernoname{subsubsection}{\secn@m.\the\subsecno\the\subsubsecno\the\subsubsubsecno}%
{\secn@m.\the\subsecno.\the\subsubsecno.\the\subsubsubsecno.} #1}%
\par\nobreak\medskip\nobreak\noindent\ignorespaces}


\def\newnewsec#1#2\par{\global\advance\secno by1%
{\toks0{#2}\message{(\secn@m. \the\toks0)}}%
\global\subsecno=0\global\subsubsecno=0\eqnres@t\let\s@csym\secsym\xdef\secn@m{\the\secno}\noindent
\ifnum\lastpenalty>9000\else\bigbreak\fi
\noindent{\bf\hyperdef\hypernoname{section}{\secn@m}{\secn@m.} #2}%
\writetoca{{\string\hyperref{}{section}{\the\secno}{\bf \the\secno\quad}} {\bf #2}}
\DefWarn#1%
\xdef#1{\noexpand\hyperref{}{section}{\the\secno}%
{\the\secno}}\writedef{#1\leftbracket#1}\wrlabeL{#1=#1}%
\par\nobreak\medskip\nobreak\noindent\ignorespaces}

\def\newsubsec#1#2\par{\global\advance\subsecno by1%
{\toks0{#2}\message{(\secn@m.\the\subsecno. \the\toks0)}}%
\global\subsubsecno=0%
\ifnum\lastpenalty>9000\else\bigbreak\fi
\noindent{\it\hyperdef\hypernoname{subsection}{\secn@m.\the\subsecno}%
{\secn@m.\the\subsecno.} #2}
\DefWarn#1%
\xdef#1{\noexpand\hyperref{}{subsection}{\secn@m.\the\subsecno}%
{\secn@m.\the\subsecno}}\writedef{#1\leftbracket#1}\wrlabeL{#1=#1}%
\writetoca{\string\hskip1.45cm
{\string\hyperref{}{subsection}{\secn@m.\the\subsecno}{\secn@m.\the\subsecno.}}
{#2}}%
\par\nobreak\medskip\nobreak\noindent\ignorespaces}

\def\newsubsecstar#1#2\par{\global\advance\subsecno by1%
{\toks0{#2}\message{(\secn@m.\the\subsecno. \the\toks0)}}%
\global\subsubsecno=0%
\ifnum\lastpenalty>9000\else\bigbreak\fi
\noindent{\it\hyperdef\hypernoname{subsection}{\secn@m.\the\subsecno}%
{\secn@m.\the\subsecno.} #2}
\DefWarn#1%
\xdef#1{\noexpand\hyperref{}{subsection}{\secn@m.\the\subsecno}%
{\secn@m.\the\subsecno}}\writedef{#1\leftbracket#1}\wrlabeL{#1=#1}%
\par\nobreak\medskip\nobreak\noindent\ignorespaces}

\def\newsubsubsec#1#2\par{\global\advance\subsubsecno by1%
{\toks0{#2}\message{(\secn@m.\the\subsecno.\the\subsubsecno. \the\toks0)}}%
\global\subsubsubsecno=0%
\ifnum\lastpenalty>9000\else\bigbreak\fi
\noindent{\it\hyperdef\hypernoname{subsubsection}{\secn@m.\the\subsecno.\the\subsubsecno}%
{\secn@m.\the\subsecno.\the\subsubsecno.} #2}
\DefWarn#1%
\xdef#1{\noexpand\hyperref{}{subsubsection}{\secn@m.\the\subsecno.\the\subsubsecno}%
{\secn@m.\the\subsecno.\the\subsubsecno}}\writedef{#1\leftbracket#1}\wrlabeL{#1=#1}%
\par\nobreak\medskip\nobreak\noindent\ignorespaces}

\def\newsubsubsubsec#1#2\par{\global\advance\subsubsubsecno by1%
{\toks0{#2}\message{(\secn@m.\the\subsecno.\the\subsubsecno.\the\subsubsubsecno \the\toks0)}}%
\ifnum\lastpenalty>9000\else\bigbreak\fi
\noindent{\it\hyperdef\hypernoname{subsubsection}{\secn@m.\the\subsecno\the\subsubsecno\the\subsubsubsecno}%
{\secn@m.\the\subsecno.\the\subsubsecno.\the\subsubsubsecno.} #2}
\DefWarn#1%
\xdef#1{\noexpand\hyperref{}{subsubsubsection}{\secn@m.\the\subsecno.\the\subsubsecno.\the\subsubsubsecno}%
{\secn@m.\the\subsecno.\the\subsubsecno.\the\subsubsubsecno}}\writedef{#1\leftbracket#1}\wrlabeL{#1=#1}%
\par\nobreak\medskip\nobreak\noindent\ignorespaces}

\def\appendix#1#2{\global\meqno=1\global\subsecno=0\global\subsubsecno=0\xdef\secsym{\hbox{#1.}}%
\bigbreak\bigskip\noindent{\bf Appendix \hyperdef\hypernoname{appendix}{#1}%
{#1.} #2}{\toks0{(#1. #2)}\message{\the\toks0}}%
\xdef\s@csym{#1.}\xdef\secn@m{#1}%
\writetoca{{\string\hyperref{}{appendix}{#1}{\bf {#1}\quad}} {\bf #2}}%
\par\nobreak\medskip\nobreak}

%
\def\checkm@de#1#2{\ifmmode{\def\f@rst##1{##1}\hyperdef\hypernoname{equation}%
{#1}{#2}}\else\hyperref{}{equation}{#1}{#2}\fi}
\def\eqnn#1{\DefWarn#1\xdef #1{(\noexpand\relax\noexpand\checkm@de%
{\s@csym\the\meqno}{\secsym\the\meqno})}%
\wrlabeL#1\writedef{#1\leftbracket#1}\global\advance\meqno by1}
\def\f@rst#1{\c@t#1a\em@ark}\def\c@t#1#2\em@ark{#1}
\def\eqna#1{\DefWarn#1\wrlabeL{#1$\{\}$}%
\xdef #1##1{(\noexpand\relax\noexpand\checkm@de%
{\s@csym\the\meqno\noexpand\f@rst{##1}1}{\hbox{$\secsym\the\meqno##1$}})}
\writedef{#1\numbersign1\leftbracket#1{\numbersign1}}\global\advance\meqno by1}
\def\eqn#1#2{\DefWarn#1%
\xdef #1{(\noexpand\hyperref{}{equation}{\s@csym\the\meqno}%
{\secsym\the\meqno})}$$#2\eqno(\hyperdef\hypernoname{equation}%
{\s@csym\the\meqno}{\secsym\the\meqno})\eqlabeL#1$$%
\writedef{#1\leftbracket#1}\global\advance\meqno by1}
\def\xeqn{\expandafter\xe@n}\def\xe@n(#1){#1}
\def\xeqna#1{\expandafter\xe@n#1}
\def\eqns#1{(\e@ns #1{\hbox{}})}
\def\e@ns#1{\ifx\UNd@FiNeD#1\message{eqnlabel \string#1 is undefined.}%
\xdef#1{(?.?)}\fi{\let\hyperref=\relax\xdef\next{#1}}%
\ifx\next\em@rk\def\next{}\else%
\ifx\next#1\xeqn#1\else\def\n@xt{#1}\ifx\n@xt\next#1\else\xeqna#1\fi
\fi\let\next=\e@ns\fi\next}
\def\etag#1{\eqnn#1\eqno#1}\def\etaga#1{\eqna#1\eqno#1}
\def\DefWarn#1{}
%
\newskip\footskip\footskip14pt plus 1pt minus 1pt 
\def\footnotefont{\ninepoint}\def\f@t#1{\footnotefont #1\@foot}
\def\f@@t{\baselineskip\footskip\bgroup\footnotefont\aftergroup\@foot\let\next}
\setbox\strutbox=\hbox{\vrule height9.5pt depth4.5pt width0pt}
\global\newcount\ftno \global\ftno=0
\def\foot{\global\advance\ftno by1\def\foot@rg{\hyperref{}{footnote}%
{\the\ftno}{\the\ftno}\xdef\foot@rg{\noexpand\hyperdef\noexpand\hypernoname%
{footnote}{\the\ftno}{\the\ftno}}}\footnote{$^{\foot@rg}$}}
%
%
%
\global\newcount\refno \global\refno=1
\newwrite\rfile
\def\ref{[\hyperref{}{reference}{\the\refno}{\the\refno}]\nref}
\def\nref#1{\DefWarn#1%
\xdef#1{[\noexpand\hyperref{}{reference}{\the\refno}{\the\refno}]}%
\writedef{#1\leftbracket#1}%
\ifnum\refno=1\immediate\openout\rfile=\jobname.refs\fi
\chardef\wfile=\rfile\immediate\write\rfile{\noexpand\item{[\noexpand\hyperdef%
\noexpand\hypernoname{reference}{\the\refno}{\the\refno}]\ }%
\reflabeL{#1\hskip.31in}\pctsign}\global\advance\refno by1\findarg}
\def\findarg#1#{\begingroup\obeylines\newlinechar=`\^^M\pass@rg}
{\obeylines\gdef\pass@rg#1{\writ@line\relax #1^^M\hbox{}^^M}%
\gdef\writ@line#1^^M{\expandafter\toks0\expandafter{\striprel@x #1}%
\edef\next{\the\toks0}\ifx\next\em@rk\let\next=\endgroup\else\ifx\next\empty%
\else\immediate\write\wfile{\the\toks0}\fi\let\next=\writ@line\fi\next\relax}}
\def\striprel@x#1{} \def\em@rk{\hbox{}}
\def\lref{\begingroup\obeylines\lr@f}
\def\lr@f#1#2{\DefWarn#1\gdef#1{\let#1=\UNd@FiNeD\ref#1{#2}}\endgroup\unskip}
\def\semi{;\hfil\break}
\def\addref#1{\immediate\write\rfile{\noexpand\item{}#1}} 
\def\listrefs{\vfill\supereject\immediate\closeout\rfile\writestoppt
\baselineskip=\footskip\centerline{{\bf References}}\bigskip{\parindent=20pt%
\frenchspacing\escapechar=` \input \jobname.refs\vfill\eject}\nonfrenchspacing}
\def\startrefs#1{\immediate\openout\rfile=\jobname.refs\refno=#1}
\def\xref{\expandafter\xr@f}\def\xr@f[#1]{#1}
\def\refs#1{\count255=1[\r@fs #1{\hbox{}}]}
\def\r@fs#1{\ifx\UNd@FiNeD#1\message{reflabel \string#1 is undefined.}%
\nref#1{need to supply reference \string#1.}\fi%
\vphantom{\hphantom{#1}}{\let\hyperref=\relax\xdef\next{#1}}%
\ifx\next\em@rk\def\next{}%
\else\ifx\next#1\ifodd\count255\relax\xref#1\count255=0\fi%
\else#1\count255=1\fi\let\next=\r@fs\fi\next}
\def\figures{\centerline{{\bf Figure Captions}}\medskip\parindent=40pt%
\def\fig##1##2{\medskip\item{Fig.~\hyperdef\hypernoname{figure}{##1}{##1}.  }%
##2}}
%
\newwrite\ffile\global\newcount\figno \global\figno=1
\def\fig{fig.~\hyperref{}{figure}{\the\figno}{\the\figno}\nfig}
\def\nfig#1{\DefWarn#1%
\xdef#1{fig.~\noexpand\hyperref{}{figure}{\the\figno}{\the\figno}}%
\writedef{#1\leftbracket fig.\noexpand~\xfig#1}%
\ifnum\figno=1\immediate\openout\ffile=\jobname.figs\fi\chardef\wfile=\ffile%
{\let\hyperref=\relax
\immediate\write\ffile{\noexpand\medskip\noexpand\item{Fig.\ %
\noexpand\hyperdef\noexpand\hypernoname{figure}{\the\figno}{\the\figno}. }
\reflabeL{#1\hskip.55in}\pctsign}}\global\advance\figno by1\findarg}
\def\xfig{\expandafter\xf@g}\def\xf@g fig.\penalty\@M\ {}
\def\figs#1{figs.~\f@gs #1{\hbox{}}}
\def\f@gs#1{{\let\hyperref=\relax\xdef\next{#1}}\ifx\next\em@rk\def\next{}\else
\ifx\next#1\xfig #1\else#1\fi\let\next=\f@gs\fi\next}
%
\def\figin{\epsfcheck\figin}\def\figins{\epsfcheck\figins}
\def\epsfcheck{\ifx\epsfbox\UnDeFiNeD
\message{(NO epsf.tex, FIGURES WILL BE IGNORED)}
\gdef\figin##1{\vskip2in}\gdef\figins##1{\hskip.5in}
\else\message{(FIGURES WILL BE INCLUDED)}%
\gdef\figin##1{##1}\gdef\figins##1{##1}\fi}
\def\figinsert{\goodbreak\topinsert}
\def\ifig#1#2#3{\DefWarn#1\xdef#1{fig.~\the\figno}
\writedef{#1\leftbracket fig.\noexpand~\the\figno}%
\figinsert\figin{\centerline{#3}}
\smallskip
\leftskip=0pt \rightskip=0pt
\baselineskip12pt\noindent
{{\bf Fig.~\the\figno}\ \ninepoint #2}
\medskip
\global\advance\figno by1\par\endinsert}
\newwrite\lfile
{\escapechar-1\xdef\pctsign{\string\%}\xdef\leftbracket{\string\{}
\xdef\rightbracket{\string\}}\xdef\numbersign{\string\#}}
\def\writedefs{\immediate\openout\lfile=label.defs \def\writedef##1{%
{\let\hyperref=\relax\let\hyperdef=\relax\let\hypernoname=\relax
 \immediate\write\lfile{\string\checkdef\string##1\rightbracket}}}}%
\def\writestop{\def\writestoppt{\immediate\write\lfile{\string\pageno
 \the\pageno\string\startrefs\leftbracket\the\refno\rightbracket
 \string\def\string\secsym\leftbracket\secsym\rightbracket
 \string\secno\the\secno\string\meqno\the\meqno}\immediate\closeout\lfile}}
\def\writestoppt{}\def\writedef#1{}

\def\seclab#1\par{\DefWarn#1%
\xdef #1{\noexpand\hyperref{}{section}{\the\secno}{\the\secno}}%
\writedef{#1\leftbracket#1}\wrlabeL{#1=#1}\par%
\nobreak\medskip\nobreak\noindent\ignorespaces}
\def\subseclab#1\par{\DefWarn#1%
\xdef #1{\noexpand\hyperref{}{subsection}{\the\secno.\the\subsecno}%
{\the\secno.\the\subsecno}}\writedef{#1\leftbracket#1}\wrlabeL{#1=#1}\par%
\nobreak\medskip\nobreak\noindent\ignorespaces}
\def\subsubseclab#1\par{\DefWarn#1%
\xdef#1{\noexpand\hyperref{}{subsubsection}{\the\secno.\the\subsecno.\the\subsubsecno}%
{\the\secno.\the\subsecno.\the\subsubsecno}}\writedef{#1\leftbracket#1}\wrlabeL{#1=#1}\par%
\nobreak\medskip\nobreak\noindent\ignorespaces}
\def\applab#1\par{\DefWarn#1%
\xdef#1{\noexpand\hyperref{}{appendix}{\secn@m}{\secn@m}}%
\writedef{#1\leftbracket#1}\wrlabeL{#1=#1}%
\par\nobreak\medskip\nobreak\noindent\ignorespaces}
\def\appsublab#1{\DefWarn#1%
\xdef #1{\noexpand\hyperref{}{appendix}{\secn@m.\the\subsecno}{\secn@m.\the\subsecno}}%
\writedef{#1\leftbracket#1}\wrlabeL{#1=#1}}
\newwrite\tfile \def\writetoca#1{}
\def\leaderfill{\leaders\hbox to 1em{\hss.\hss}\hfill}
\def\writetoc{\immediate\openout\tfile=\jobname.toc
   \def\writetoca##1{{\edef\next{\write\tfile{\noindent ##1
   \string\leaderfill{
   \string\hyperref{}{page}{\noexpand\number\pageno}%
   {\noexpand\number\pageno}} \par}}\next}}
}
\newread\ch@ckfile
\def\listtoc{\immediate\closeout\tfile\immediate\openin\ch@ckfile=\jobname.toc
\ifeof\ch@ckfile\message{no file \jobname.toc, no table of contents this pass}%
\else\closein\ch@ckfile\centerline{\bf Contents}\nobreak\medskip%
{\baselineskip=15.5pt\footnotefont\parskip=0pt\catcode`\@=11\input\jobname.toc
\catcode`\@=12\bigbreak\bigskip}\fi}
\catcode`\@=12 
\font\authorfont=cmcsc10
\def\tenpoint{\def\rm{\fam0\tenrm}
\textfont0=\tenrm \scriptfont0=\sevenrm \scriptscriptfont0=\fiverm
\textfont1=\teni  \scriptfont1=\seveni  \scriptscriptfont1=\fivei
\textfont2=\tensy \scriptfont2=\sevensy \scriptscriptfont2=\fivesy
\textfont\itfam=\tenit \def\it{\fam\itfam\tenit}\def\footnotefont{\ninepoint}%
\textfont\bffam=\tenbf \def\bf{\fam\bffam\tenbf}\def\sl{\fam\slfam\tensl}\rm}
\font\ninerm=cmr9 \font\sixrm=cmr6 \font\ninei=cmmi9 \font\sixi=cmmi6
\font\ninesy=cmsy9 \font\sixsy=cmsy6 \font\ninebf=cmbx9
\font\nineit=cmti9 \font\ninesl=cmsl9 \skewchar\ninei='177
\skewchar\sixi='177 \skewchar\ninesy='60 \skewchar\sixsy='60
\def\ninepoint{\def\rm{\fam0\ninerm}
\textfont0=\ninerm \scriptfont0=\sixrm \scriptscriptfont0=\fiverm
\textfont1=\ninei \scriptfont1=\sixi \scriptscriptfont1=\fivei
\textfont2=\ninesy \scriptfont2=\sixsy \scriptscriptfont2=\fivesy
\textfont\itfam=\ninei \def\it{\fam\itfam\nineit}\def\sl{\fam\slfam\ninesl}%
\textfont\bffam=\ninebf \def\bf{\fam\bffam\ninebf}\rm}
%
\hyphenation{anom-aly anom-alies coun-ter-term coun-ter-terms}

\def\tikzcaption#1#2{\DefWarn#1\xdef#1{Fig.~\the\figno}
\writedef{#1\leftbracket Fig.\noexpand~\the\figno}%
{
\smallskip
\leftskip=20pt \rightskip=20pt \baselineskip12pt\noindent
{{\bf Fig.~\the\figno}\ \ninepoint #2}
\bigskip
\global\advance\figno by1 \par}}

\def\ntoalpha#1{%
\ifcase#1%
@%
\or A\or B\or C\or D\or E\or F\or G\or H\or I\or J\or K\or L\or M%
\fi
}

\global\newcount\appno \global\appno=1
\def\applab#1{\xdef #1{\ntoalpha{\appno}}\writedef{#1\leftbracket#1}\wrlabeL{#1=#1}
\global\advance\appno by1}

\def\preprint#1 #2\par{\rightline{\vbox{\baselineskip12pt\hbox{#1}\hbox{#2}}}\vskip2cm}
%
\def\title#1\par{\centerline{\bf #1}\nopagenumbers\pageno=0}
\def\author#1\par{\bigskip\bigskip\centerline{#1}}

\newcount\addressno

\def\email#1#2{
\footnote{\null}{\kern-\parindent \llap{$^#1$\hskip1pt}email: #2}}

\def\startcenter{%
  \par
  \begingroup
  \leftskip=0pt plus 1fil
  \rightskip=\leftskip
  \parindent=0pt
  \parfillskip=0pt
}
\def\stopcenter{\endgroup}

\def\address{\bigskip%
  \ifnum\the\addressno=0\else\stopcenter\endgroup\fi
  \advance\addressno by 1%
  \begingroup
  \startcenter
  \it
  \obeylines
  \addressAux
}
\def\addressAux#1{#1}

\def\abstract{\stopcenter\endgroup\bigskip\bigskip\noindent}

\def\Dsl{\,\raise.15ex\hbox{/}\mkern-13.5mu D} 
\def\dsl{\raise.15ex\hbox{/}\kern-.57em\partial}
\def\tr{{\rm tr}} \def\Tr{{\rm Tr}}
\def\boxeqn#1{\vcenter{\vbox{\hrule\hbox{\vrule\kern3pt\vbox{\kern3pt
	\hbox{${\displaystyle #1}$}\kern3pt}\kern3pt\vrule}\hrule}}}
\def\grad#1{\,\nabla\!_{{#1}}\,}
\def\gradgrad#1#2{\,\nabla\!_{{#1}}\nabla\!_{{#2}}\,}
\def\lform{\hbox{$\sqcup$}\llap{\hbox{$\sqcap$}}}
\def\lie{\hbox{\it\$}} 

\def\ap{{\alpha^{\prime}}}
\def\halfap#1{\Big({\ap\over 2}\Big)^{\mkern-4mu #1}}
\def\a{\alpha}
\def\b{{\beta}}
\def\g{{\gamma}}
\def\d{{\delta}}
\def\e{{\epsilon}}
\def\l{\lambda}
\def\k{{\kappa}}
\def\s{{\sigma}}
\def\t{{\theta}}
\def\om{{\omega}}
\def\lb{{\overline\lambda}}
\def\llb{(\l\lb)}
\def\wb{{\overline w}}
\def\half{{1\over 2}}
\def\p{{\partial}}
\def\pb{{\overline\partial}}
\def\tb{{\overline\theta}}
\def\bar{\overline}
\def\({\left(}
\def\){\right)}
\def\dz{{\rm d}z}

\def\cA{{\cal A}}
\def\cF{{\cal F}}
\def\cJ{{\cal J}}
\def\cK{{\cal K}}
\def\cI{{\cal I}}
\def\cV{{\cal V}}
\def\cW{{\cal W}}
\def\cY{{\cal Y}}
\def\cZ{{\cal Z}}

\def\bV{{\Bbb V}}
\def\bA{{\Bbb A}}
\def\bW{{\Bbb W}}
\def\bF{{\Bbb F}}

\def\Box{\square}
\def\AYM{A^{\rm SYM}}

\def\psum{\mathop{\sum\nolimits'}}

\def\len#1{{%
\def\Dlen{\left|\mkern-1mu #1\mkern -0.5mu\right|}%
\def\Sslen{\left|\mkern-1.3mu #1\mkern -1.3mu\right|}%
\def\SSlen{\left|\mkern-2.8mu #1\mkern-1.3mu\right|}%
\mathchoice{\Dlen}{\Dlen}{\Sslen}{\SSlen}}}

\def\perm#1{{\rm perm}#1}
\def\eikx{{\bigl\langle \prod_{j=1}^4 {\rm e}^{i k^j\cdot x^j}\bigr\rangle}}
\def\ImOmega{\Im\Omega}
\def\lalb#1{(\l\g^{#1}\lb)}
\def\Im{\mathop{{\rm Im}}} 
\def\sfrac#1/#2{\kern.1em\raise.5ex\hbox{\the\scriptfont0 #1}%
\kern-.1em/\kern-.15em\lower.25ex\hbox{\the\scriptfont0 #2}}

\font\tenshuffle=shuffle10 \font\sevenshuffle=shuffle7 \font\fiveshuffle=shuffle7 at 5pt
\def\shuffle{{%
\def\Dshuffle{\mathbin{\hbox{\tenshuffle\char'001}}}%
\def\Sshuffle{\mathbin{\hbox{\sevenshuffle\char'001}}}%
\def\SSshuffle{\mathbin{\hbox{\fiveshuffle\char'001}}}%
\mathchoice{\Dshuffle}{\Dshuffle}{\Sshuffle}{\SSshuffle}}}

\font\tenwedge=stmary10 \font\sevenwedge=stmary7 \font\fivewedge=stmary5
\def\owedge{%
\def\Dowedge{\mathbin{\hbox{\tenwedge\char"3F}}}%
\def\Sowedge{\mathbin{\hbox{\sevenwedge\char"3F}}}%
\def\SSowedge{\mathbin{\hbox{\fivewedge\char"3F}}}%
\mathchoice{\Dowedge}{\Dowedge}{\Sowedge}{\SSowedge}}

\def\qed{\hbox{\hskip 3pt
\vbox{\hrule\hbox to 7pt{\vrule height 7pt\hfill\vrule}
\hrule}}\hskip3pt}

\overfullrule=0pt\relax

\frenchspacing

\def\checkdef#1#2{
\ifx\UndeFined#1%
	\def#1{#2}
\else
	\immediate\write16{*** BUG ***: the label \string#1 is already defined ***}
\fi
}
\newread\instream
\def\openlabels{
\openin\instream= label.defs
\ifeof\instream\message{No labels in advance yet. Wait till next pass.}
\else\closein\instream \input label.defs
\fi}

\openin\instream= label.defs
\ifeof\instream\message{No labels in advance yet. Wait till next pass.}
\else\closein\instream \input label.defs
\fi
\writedefs

\def\arXiv:#1].{\hepthStrip#1 \nil}
\def\hepthStrip#1 #2\nil{\href{http://arxiv.org/abs/#1}{arXiv:#1 #2\unskip}].}

\font\frakfont=eufm10 at 10pt
\def\ce{\mathord{\hbox{\frakfont e}}}
\def\cf{\mathord{\hbox{\frakfont f}}}
\def\cm{\mathord{\hbox{\frakfont M}}}

\input amssym

\def\textbf#1{{\bf #1}}
\def\paragraph#1{\medskip\noindent{\it #1.}}
\def\frac#1#2{{#1\over #2}}
\def\KN{{\rm KN}}
\def\qeq{\buildrel ? \over =}
\def\V#1{V_{#1}^{(w_#1)}}
\def\Vz#1{V_{#1}^{(0)}}
\def\Vo#1{V_{#1}^{(1)}}
\def\Vt#1{V_{#1}^{(2)}}
\def\Vi{V_{i}^{(w_i)}}
\def\Vj{V_{j}^{(w_j)}}
\def\Vk{V_{k}^{(w_k)}}

\title Manifest M\"obius invariance of massive tree-level three-point amplitudes

\title in pure spinor superspace

\author
Chen Huang\email{a}{chen.huang@soton.ac.uk}$^a$,
Carlos R. Mafra\email{b}{c.r.mafra@soton.ac.uk}$^b$ and
Yi-Xiao Tao\email{c}{taoyx21@mails.tsinghua.edu.cn}$^c$

\address
$^{a,b}$Mathematical Sciences and STAG Research Centre, University of Southampton,
Highfield, Southampton, SO17 1BJ, UK

\address
$^c$Department of Mathematical Sciences, Tsinghua University,
Beijing 100084, China

\address
$^c$Nordita, KTH Royal Institute of Technology and Stockholm University,
Hannes Alf\'{v}ens v\"{a}g 12, SE-106 91 Stockholm, Sweden

\abstract
Using
BRST cohomology properties in pure spinor superspace and identities for OPE brackets of
non-free fields, we obtain a new compact nested-bracket representation of massive
tree-level three-point open-string
amplitudes in which M\"obius invariance is manifest. Explicit
superspace calculations for amplitudes with
level-one massive states confirm this finding,
and we derive new BRST recurrence relations among three-point numerators
to
extend the result to arbitrary mass levels. This provides a manifestly
M\"obius-invariant expression for massive three-point amplitudes in the pure spinor formalism.

\Date{March 2026}

\lref\openstrings{
	C.~Angelantonj and A.~Sagnotti,
	``Open strings,''
	Phys. Rept. \textbf{371}, 1-150 (2002)
	[erratum: Phys. Rept. \textbf{376}, no.6, 407 (2003)]
	[arXiv:hep-th/0204089 [hep-th]].
}
\lref\schwarzrev{
	J.~H.~Schwarz,
	``Superstring Theory,''
	Phys. Rept. \textbf{89}, 223-322 (1982)
}

\lref\excursion{
	C.~Markou and E.~Skvortsov,
	``An excursion into the string spectrum,''
	JHEP \textbf{12}, 055 (2023)
	[arXiv:2309.15988 [hep-th]].
}

\lref\Vcolor{
	T.~van Ritbergen, A.~N.~Schellekens and J.~A.~M.~Vermaseren,
  	``Group theory factors for Feynman diagrams,''
	Int.\ J.\ Mod.\ Phys.\ A {\bf 14}, 41 (1999).
	[hep-ph/9802376].
}
\lref\eulerian{
	R.~Bandiera and C.~R.~Mafra,
	``A closed-formula solution to the color-trace decomposition problem,''
	[arXiv:2009.02534 [math.CO]].
}

\lref\PSthreemass{
	S.~Chakrabarti, S.~P.~Kashyap and M.~Verma,
	``Amplitudes Involving Massive States Using Pure Spinor Formalism,''
	JHEP \textbf{12}, 071 (2018)
	[arXiv:1808.08735 [hep-th]].
}

\lref\fpointbos{
	M.~Firrotta,
	``Veneziano and Shapiro-Virasoro amplitudes of arbitrarily excited strings,''
	JHEP \textbf{06}, 115 (2024)
	[arXiv:2402.16183 [hep-th]].
}

\lref\SiegelYI{
	W.Siegel,
	``Superfields in Higher Dimensional Space-time,''
	Phys. Lett.B {\bf 80}, 220~(1979)
}
\lref\siegel{
	W.~Siegel,
	``Classical Superstring Mechanics,''
	Nucl.\ Phys.\  {\bf B263}, 93 (1986).
}

\lref\grosschaos{
	D.~J.~Gross and V.~Rosenhaus,
	``Chaotic scattering of highly excited strings,''
	JHEP \textbf{05}, 048 (2021)
	[arXiv:2103.15301 [hep-th]].
}

\lref\thielemans{
	K.~Thielemans,
	``An Algorithmic approach to operator product expansions, W algebras and W strings,''
	[arXiv:hep-th/9506159 [hep-th]].
}

\lref\olimass{
	O.~Schlotterer,
	``Higher Spin Scattering in Superstring Theory,''
	Nucl. Phys. B \textbf{849}, 433-460 (2011)
	[arXiv:1011.1235 [hep-th]].
}

\lref\hSYM{
	C.~R.~Mafra and O.~Schlotterer,
	``Solution to the nonlinear field equations of ten dimensional supersymmetric Yang-Mills theory,''
	Phys. Rev. D \textbf{92}, no.6, 066001 (2015)
	[arXiv:1501.05562 [hep-th]].
}

\lref\stiet{
	D.~Lust, C.~Markou, P.~Mazloumi and S.~Stieberger,
	``Extracting bigravity from string theory,''
	JHEP \textbf{12}, 220 (2021)
	[arXiv:2106.04614 [hep-th]].
}
\lref\bigpicture{
	N.~Berkovits, M.~T.~Hatsuda and W.~Siegel,
	``The Big picture,''
	Nucl. Phys. B \textbf{371}, 434-466 (1992)
	[arXiv:hep-th/9108021 [hep-th]].
}
\lref\bianchi{
	M.~Bianchi and A.~L.~Guerrieri,
	``On the soft limit of open string disk amplitudes with massive states,''
	JHEP \textbf{09}, 164 (2015)
	[arXiv:1505.05854 [hep-th]].
}

\lref\intUpaper{
	S.~Chakrabarti, S.~P.~Kashyap and M.~Verma,
	``Integrated Massive Vertex Operator in Pure Spinor Formalism,''
	JHEP \textbf{10}, 147 (2018)
	[arXiv:1802.04486 [hep-th]].
}

\lref\massivevone{
	S.P.~Kashyap, C.R.~Mafra, M.~Verma and L.A.~Ypanaque,
	``A relation between massive and massless string tree amplitudes,''
	[arXiv:2311.12100 [hep-th]].
}
\lref\nptMethod{
	C.~R.~Mafra, O.~Schlotterer, S.~Stieberger and D.~Tsimpis,
	``A recursive method for SYM n-point tree amplitudes,''
	Phys.\ Rev.\ D {\bf 83}, 126012 (2011).
	[arXiv:1012.3981 [hep-th]].
}
\lref\towardsFT{
	C.R.~Mafra,
	``Towards Field Theory Amplitudes From the Cohomology of Pure Spinor Superspace,''
	JHEP {\bf 1011}, 096 (2010).
	[arXiv:1007.3639 [hep-th]].
}
\lref\cdescent{
	C.R.~Mafra,
	``KK-like relations of $\alpha$' corrections to disk amplitudes,''
	JHEP \textbf{03}, 012 (2022)
	[arXiv:2108.01081 [hep-th]].
}
\lref\massSweden{
       	M.~Guillen, H.~Johansson, R.~L.~Jusinskas and O.~Schlotterer,
	``Scattering Massive String Resonances through Field-Theory Methods,''
	Phys. Rev. Lett. \textbf{127}, no.5, 051601 (2021)
	[arXiv:2104.03314 [hep-th]].
}
\lref\BerendsME{
	F.A.~Berends and W.T.~Giele,
	``Recursive Calculations for Processes with n Gluons,''
	Nucl.\ Phys.\ B {\bf 306}, 759 (1988).
}

\lref\MSSI{
	C.R.~Mafra, O.~Schlotterer and S.~Stieberger,
	``Complete N-Point Superstring Disk Amplitude I. Pure Spinor Computation,''
	Nucl.\ Phys.\ B {\bf 873}, 419 (2013).
	[arXiv:1106.2645 [hep-th]].
}
\lref\MSSII{
	C.~R.~Mafra, O.~Schlotterer and S.~Stieberger,
	``Complete N-Point Superstring Disk Amplitude II. Amplitude
	and Hypergeometric Function Structure,''
	Nucl.\ Phys.\ B {\bf 873}, 461 (2013).
	[arXiv:1106.2646 [hep-th]].
}
\lref\PSthreemass{
	S.~Chakrabarti, S.~P.~Kashyap and M.~Verma,
	``Amplitudes Involving Massive States Using Pure Spinor Formalism,''
	JHEP \textbf{12}, 071 (2018)
	[arXiv:1808.08735 [hep-th]].
}
\lref\masstheta{
	S.~Chakrabarti, S.~P.~Kashyap and M.~Verma,
	``Theta Expansion of First Massive Vertex Operator in Pure Spinor,''
	JHEP \textbf{01}, 019 (2018)
	[arXiv:1706.01196 [hep-th]].
}
\lref\PSS{
	C.R.~Mafra,
	``PSS: A FORM Program to Evaluate Pure Spinor Superspace Expressions,''
	[arXiv:1007.4999 [hep-th]].
}

\lref\ICTP{
	N.~Berkovits,
  	``ICTP lectures on covariant quantization of the superstring,''
	[hep-th/0209059].
}

\lref\BCpaper{
	N.~Berkovits and O.~Chandia,
	``Massive superstring vertex operator in D = 10 superspace,''
	JHEP \textbf{08}, 040 (2002)
	[arXiv:hep-th/0204121 [hep-th]].
}

\lref\psf{
 	N.~Berkovits,
	``Super-Poincare covariant quantization of the superstring,''
	JHEP {\bf 0004}, 018 (2000)
	[arXiv:hep-th/0001035].
}

\lref\wittentwistor{
	E.Witten,
        ``Twistor-Like Transform In Ten-Dimensions''
        Nucl.Phys. B {\bf 266}, 245~(1986)
}
\lref\higherSYM{
	C.R.~Mafra and O.~Schlotterer,
	``A solution to the non-linear equations of D=10 super Yang-Mills theory,''
	Phys.\ Rev.\ D {\bf 92}, no. 6, 066001 (2015).
	[arXiv:1501.05562 [hep-th]].
}
\lref\treereview{
	C.~R.~Mafra and O.~Schlotterer,
	``Tree-level amplitudes from the pure spinor superstring,''
	Phys. Rept. \textbf{1020}, 1-162 (2023)
	[arXiv:2210.14241 [hep-th]].
}
\lref\FORM{
	J.A.M.~Vermaseren,
	``New features of FORM,''
	arXiv:math-ph/0010025.
\semi
	M.~Tentyukov and J.A.M.~Vermaseren,
	``The multithreaded version of FORM,''
	arXiv:hep-ph/0702279.
}
\lref\PSspace{
	N.~Berkovits,
	``Explaining Pure Spinor Superspace,''
	[arXiv:hep-th/0612021 [hep-th]].
}
\lref\UV{
	S.P.~Kashyap, C.R.~Mafra, M.~Verma and L.~Ypanaqu\'e,
	``Massless representation of massive superfields and tree amplitudes with the pure spinor formalism,''
	[arXiv:2407.02436 [hep-th]].
}
\lref\farril{
	J.~M.~Figueroa-O'Farrill,
	``N=2 structures in all string theories,''
	J. Math. Phys. \textbf{38}, 5559-5575 (1997)
	[arXiv:hep-th/9507145 [hep-th]].
}
\lref\knuthconcrete{
	R. Graham, D.E. Knuth, and O. Patashnik,
 	``Concrete Mathematics: A Foundation for Computer Science'',
	Addison-Wesley Longman Publishing Co., Inc.,
	Boston, MA, USA, (1994).
}

\lref\leerev{
	J.~C.~Lee and Y.~Yang,
	``Review on high energy string scattering amplitudes and symmetries of string theory,''
	Phys. Rept. \textbf{1142}, 1-203 (2025)
}
\lref\psspec{
	N.~Berkovits,
	``Cohomology in the pure spinor formalism for the superstring,''
	JHEP \textbf{09}, 046 (2000)
	[arXiv:hep-th/0006003 [hep-th]].
}
\lref\psspecR{
	N.~Berkovits and R.~Lipinski Jusinskas,
	``Light-Cone Analysis of the Pure Spinor Formalism for the Superstring,''
	JHEP \textbf{08}, 102 (2014)
	[arXiv:1406.2290 [hep-th]].
}
\lref\DDFrenannI{
	R.~L.~Jusinskas,
	``Spectrum generating algebra for the pure spinor superstring,''
	JHEP \textbf{10}, 022 (2014)
	[arXiv:1406.1902 [hep-th]].
}
\lref\DDFrenannII{
	R.~L.~Jusinskas,
	``On the field-antifield (a)symmetry of the pure spinor superstring,''
	JHEP \textbf{12}, 136 (2015)
	[arXiv:1510.05268 [hep-th]].
}

\listtoc
\writetoc
\filbreak

\newsec{Introduction}

\noindent Superstring three-point amplitudes at tree level are invariant under $SL(2,{\Bbb R})$
(M\"obius)
transformations, in fact they are constant functions on the worldsheet.
This independence is manifest in the scattering of three massless
states in the pure spinor formalism \psf\ but it is not manifest when scattering
massive states, see \refs{\grosschaos,\excursion} for emphasis on this last point. The
reason for the apparent functional dependence on the vertex positions is that the process of integrating out
the operators with non-vanishing conformal weight via OPEs
leads to various factors of $1/(z_i-z_j)^n$. They
eventually factor out and cancel against the Koba-Nielsen factor, but the cancellation only
happens after summing over
various OPE channels and using on-shell properties of the component fields as well as
momentum conservation. See
\refs{\stiet,\olimass} and \PSthreemass\ for example cancellations in the RNS and pure spinor formalisms,
and \fpointbos\ for related considerations in four-point scattering amplitudes in the bosonic string.

In this paper, we will obtain a compact {\it prescription\/} for arbitrary massive three-point
amplitudes of mass levels $(w_1,w_2,w_3)$
using the pure spinor formalism that already incorporates these cancellations.
More precisely,
we will show by direct evaluation of the pure spinor tree-level prescription of \psf\ 
that the color-ordered amplitude is given by the compact expression ($w_1\ge1$)
\eqn\introA{
A_{w_1w_2w_3}(1,2,3)= \cases{
\langle[[\hat\V1,\hat\V2]_{w_1+w_2-w_3},\hat\V3]_{2w_3}\rangle & $w_1+w_2-w_3\ge1$ \cr
(-1)^{w+1}\langle [\hat\V2,[\hat\V1,\hat\V3]_{w_1-w_2+w_3}]_{2w_2}\rangle & $w_1-w_2+w_3\ge1$
}
}
when left written in terms of OPE brackets $[\cdot,\cdot]_n$ for poles of order $n$, where $w$
denotes the total mass level (or conformal weight) $w=w_1+w_2+w_3$.
Notice that the result \introA\ is manifestly independent on the
positions of the vertex operators; the overall Koba-Nielsen factor has already been canceled
out\foot{That is why we use $\hat\Vi$ instead of $\Vi$ in \introA; to emphasize the absence of the
Koba-Nielsen factor. See section~\opepw\ for a discussion.}. Furthermore, if the two different conditions on the weight
distributions in \introA\ are met at the same time, both pure spinor superspace expressions are
the same. In addition, the factor of $(-1)^{1+w}$ on the second line of \introA\ appears strange
but it is the result of BRST cohomology considerations and it is ultimately related to the
worldsheet parity operator for massive color-ordered amplitudes, and required to give the correct
color-dressed amplitude.

However, \introA\ is not the final answer in pure spinor superspace as
one still needs to extract the superfield expressions after evaluating the OPE brackets, see
examples in the appendix~\appZ\ for vertices with weight $0$ or $1$. That is
why we called it a {\it prescription} for massive amplitudes.

\newsec{Preliminaries}

For reviews of the pure spinor formalism, see \refs{\ICTP,\treereview}.
The pure spinor unintegrated vertex operator of mass-level $N{=}w_i$ for the $i^{\rm th}$
supermultiplet, $V^{(N)}_i$, is described
by a vertex $\hat V_i^{(w_i)}$ with
conformal weight $w_i$ at zero momentum \BCpaper
\eqn\Vhatdef{
V^{(N)}_i = [\hat V^{(w_i)}_i, e^{ik_i\cdot X}]_0,
}
ensuring that the vertex $V^{(N)}_i$ has conformal weight zero. The normal ordering bracket
$[\cdot,\cdot]_0$ will be briefly reviewed below.

The unintegrated vertex operators of the massless and first-level massive supermultiplets in the
pure spinor formalism are known and given by \refs{\psf,\BCpaper}
\eqnn\Vdefs
$$\eqalignno{
V^{(0)}_i&=[\l^\a, A^i_\a]_0 &\Vdefs\cr
V_i^{(1)}&= [\l^\a \p\t^\b B^i_{\a\b}]_0 + [\l^\a\Pi^m H^m_{i\,\a}]_0
+ 2\ap [\l^\a d_\b C_i^\b{}_\a]_0
+ \ap[\l^\a N^{mn} F^i_{\a mn}]_0\,,
}$$
and are BRST closed $QV^{(w_i)}_i=0$. The notation used above is
$[ABC]_0\equiv[A,[B,C]_0]_0$ is the left-to-right normal ordering bracket.
Recall that the BRST charge in the pure spinor formalism is
given by
\eqn\BRSTcharge{
Q=\oint dz j(z),
}
where $j(z)=[\l^\a,d_\a]_0(z)$ is the BRST current and $d_\a$ is the
Green-Schwarz constraint \refs{\psf,\siegel}. With a slight abuse of
notation, we will write the action of the BRST charge $QA$ as $[Q,A]_1$ instead of $[j,A]_1$.

The worldsheet derivatives of the unintegrated vertices \Vdefs\ can be written as
\eqnn\delVs
$$\eqalignno{
\p V^{(0)}_i &= [\p\l^\a A^i_\a]_0 + [\l^\a \p A^i_\a]_0, &\delVs \cr
\p V^{(1)}_i&=
[\p\l^\a[\p\t^\b B^i_{\a\b}]_0]_0
+ [\l^\a[\p^2\t^\b B^i_{\a\b}]_0]_0
+ [\l^\a[\p\t^\b \p B^i_{\a\b}]_0]_0 \cr&
+ [\p\l^\a[\Pi^m H^m_{i\,\a}]_0]_0
+ [\l^\a[\p\Pi^m H^m_{i\,\a}]_0]_0
+ [\l^\a[\Pi^m \p H^m_{i\,\a}]_0]_0\cr&
+ 2\ap [\p\l^\a[d_\b C_i^\b{}_\a]_0]_0
+ 2\ap [\l^\a[\p d_\b C_i^\b{}_\a]_0]_0
+ 2\ap [\l^\a[d_\b \p C_i^\b{}_\a]_0]_0\cr&
+ \ap[\p\l^\a[N^{mn} F^i_{\a mn}]_0]_0
+ \ap[\l^\a[\p N^{mn} F^i_{\a mn}]_0]_0
+ \ap[\l^\a[N^{mn} \p F^i_{\a mn}]_0]_0.
}$$
In \delVs, the worldsheet derivative of superfields $K(X,\t)$ depending on $X^m,\t^\a$ can be rewritten using the
chain rule as
$\p K(X,\t) = [\p\t^\a D_\a K]_0 + [\Pi^m \p_m K]_0$,
where $D_\a = {\p\over\p\t^\a} +\half (\g^m \t)_\a \p_m$ and $\Pi^m = \p X^m +\half(\t\g^m \p\t)$.
An alternative expression for the derivative is given by
\eqn\delV{
\p\Vj = \p\hat\Vj :e^{ik_j\cdot X}: + ik^m_j \hat\Vj :\p X^m e^{ik_j\cdot X}:\,,
}
which already takes into account the subtlety described below related to the plane waves by using
a distinct normal ordering convention denoted by the double colons.
The integrated vertex operators of the massless and first-level massive supermultiplet
are given by
\refs{\psf,\intUpaper}
\eqnn\masslessU
\eqnn\intver
$$\eqalignno{
U^{(1)}_i(z) &=
[\p\t^\a A_\a^i]_0
+ [\Pi^m A^i_m]_0
+ 2\ap[d_\a W^\a_i]_0
+ \ap [N^{mn}F_i^{mn}]_0\,, &\masslessU\cr
U^{(2)}_i(z) &=
[\Pi^m\Pi^n F^i_{mn}]_0
+ [\Pi^m d_\a F_m{}^\a]_0
+ [\Pi^m \p\t^\a G_{m\a}]_0
+ [\Pi^m N^{np} F_{mnp}]_0\cr&
+ [d_\a d_\b K^{\a\b}]_0
+ [d_\a\p\t^\b F^\a{}_\b]_0
+ [d_\a N^{mn}G^\a{}_{mn}]_0
+ [\p\t^\a\p\t^\b H_{\a\b}]_0\cr&
+ [\p\t^\a N^{mn}H_{mn\a}]_0
+ [N^{mn}N^{pq}G_{mnpq}]_0 &\intver
}$$
where $[A_\a, A_m, W^\a, F^{mn}]$ are the massless superfields describing super Yang-Mills in ten
dimensions \refs{\wittentwistor,\SiegelYI}, and the various
superfields appearing in the massive vertex were
determined in \intUpaper\ and are listed in the appendix~\tensupapp. Moreover,
we omitted the
position $z$ on the right-hand side of \masslessU\ and \intver\ for brevity. 
The equations of motion of the massless superfields are \wittentwistor\ ($\p_m = ik_m$)
\eqn\RankOneEOM{
\eqalign{
D_{\a} A_{\b} + D_\b A_\a & = \g^m_{\a\b} A_m\,,\cr
D_\a F_{mn} & = \p_{m} (\g_{n} W)_\a - \p_{n} (\g_{m} W)_\a\,,
}\qquad\eqalign{
D_\a A_m &= (\g_m W)_\a + \p_m A_\a\,,  \cr
D_\a W^{\b} &= {1\over 4}(\g^{mn})^{\phantom{m}\b}_\a F_{mn}\,,
}}
while the equations of motion of the superfields at the first mass level are \masstheta,
\eqnn\eomtheta
$$\eqalignno{
D_\a G^{mn} &= -{1\over18}\p_p(\g^{pm}H^n)_\a - {1\over18}\p_p(\g^{pn}H^m)_\a\,, &\eomtheta\cr
D_\a B_{mnp} &= -{1\over18}(\g^{mn}H^p)_\a + {\ap\over18}\p_a\p_m\Big((\g^{an}H^p)_\a
-(\g^{ap}H^n)_\a\Big) + {\rm cyc}(mnp)\,,\cr
D_\a H^m_\b &=-{9\over2}G_{mn}\g^n_{\a\b}-{3\over2}\p_a B_{bcm}\g^{abc}_{\a\b}+{1\over4}\p_a
B_{bcd}\g^{mabcd}_{\a\b}\,.
}$$
where
$G^{mn} = -{1\over144}\bigl[(D\g^m H^n) + (D\g^n H^m)\bigr]$.

As expected from integrated vertex operators,
their conformal weight is shifted by ${+}1$ in comparison with the corresponding
unintegrated vertex.
In general, the integrated and unintegrated vertices for arbitrary mass level are related by
\refs{\psf,\bigpicture}
\eqn\delVQU{
\p \Vi = QU^{(w_i+1)}_i\,,
}
where $Q$ is the pure spinor BRST charge \BRSTcharge.

\newsubsec\nonfree OPEs of non-free fields

A {\it free} field is defined as a field whose OPE with itself contains a single constant term.
The fields in the pure spinor formalism are therefore not free. In this case, to compute (nested)
OPEs of composite operators we will use the
OPE bracket formalism described in \refs{\thielemans,\farril}. The OPE of the operators $A$ and $B$
is defined as ($N$ is a finite positive integer)
\eqn\opeAB{
A(z)B(w) = \sum_{n=-\infty}^{N} {[A,B]_n(w)\over (z-w)^n}
}
and the normal ordering of $A$ and $B$ is $[A,B]_0$. The OPE brackets $[\cdot,\cdot]_n$ satisfy many
identities. We list some that are used in this paper:
\eqnn\Asix
\eqnn\Aseven
\eqnn\Aten
\eqnn\Aeleven
\eqnn\Atwelve
$$\eqalignno{
[A[BC]_0]_n &= (-1)^{ab}[B[AC]_n]_0 + [[AB]_nC]_0
+ \sum_{i=1}^{n-1}{n-1\choose i}[[AB]_{n-i}C]_i &\Asix\cr
[[AB]_0 C]_n &=\sum_{j=0}^\infty\bigl(
[A[BC]_{n+j}]_{-j} + (-1)^{ab}[B[AC]_{j+1}]_{n-j-1}
\bigr)&\Aseven\cr
[AB]_n &= (-1)^{n+{ab}}\Big([BA]_n + \sum_{i=1}^\infty (-1)^{i}{1\over i!}\p^i[BA]_{n+i}\Big) &\Aten\cr
[\p A B]_n &= (1-n)[AB]_{n-1} &\Aeleven\cr
[A\p B]_n &= \p[AB]_n + (n-1)[AB]_{n-1} &\Atwelve\cr
}$$
Furthermore, $[A,\;]_1$ is a graded
derivation over all other brackets \farril,
\eqn\deriv{
[A,[B,C]_n]_1 = [[A,B]_1,C]_n + (-1)^{ab}[B,[A,C]_1]_n\,,
}
where $a,b=0,1$ depending on the Gra\ss mann nature of the operators $A,B$.
Note that if $A^{(w_A)}$ ($B^{(w_B)}$) has conformal weight $w_A$ ($w_B$) then it trivially follows that
\eqn\vanB{
[A^{(w_A)},B^{(w_B)}]_n = 0,\quad n>w_A+w_B\,.
}
To see this, it suffices to note that the bracket $[\cdot,\cdot]_n$ has conformal weight $-n$ \farril,
which would imply
a negative conformal weight if the RHS of \vanB\ were not zero.

See \refs{\thielemans,\farril,\UV} for a more complete list of
identities obeyed by the OPE brackets $[\cdot,\cdot]_n$ and the appendix~\psopeapp\ for a list of OPEs in
the pure spinor formalism.

\newsubsec\brstsec On BRST cohomology

If $A$ and $B$ are BRST closed,
$QA=QB=0$, 
then also $[A,B]_n$ is BRST closed for every $n$, $Q[A,B]_n=0$ \farril.
Moreover, if either $A$ or $B$ is BRST-exact,
then also $[A,B]_n$ is BRST exact, for every $n$. To see this, consider
$A$ to be BRST-exact and $B$ BRST-closed; $A=[Q,\Omega]_1$, $[Q,B]_1 = 0$. Then
$[Q,[\Omega,B]_n]_1 = [[Q,\Omega]_1,B]_n+(-1)^{|\Omega|}[\Omega,[Q,B]_1]_n =[A,B]_n$, where
$|\Omega|=1$ if $\Omega$ is fermionic and $0$ if bosonic.
Therefore, as
the unintegrated vertices $V^{(w_i)}_i$ are BRST closed, we conclude
\eqn\bexacs{
\langle [Q\Omega, \Vi]_n \rangle = 0,\quad \forall n\,.
}
Finally, BRST-exact terms $K=Q\Omega$ of ghost-number three in pure spinor superspace
will be indicated by $K \sim 0$. The motivation for this notation stems from the vanishing of
their pure spinor bracket \treereview;
$K\sim 0\Longrightarrow \langle K \rangle =0$.

\newsubsec\opepw OPE bracket with plane waves

It is important to note that the OPE bracket formalism for non-free fields 
can only be applied when the OPEs are characterized by a Laurent series. This excludes the
OPEs of plane wave factors since the $XX$ OPE is logarithmic.
Therefore when computing OPEs that appear in scattering amplitudes using the bracket formalism,
it is essential to treat the plane
waves separately from the OPEs of the other fields. The OPE of the plane waves is computed via
other methods and gives
an overall Koba-Nielsen factor. In practical terms, this means decomposing the vertex operators
using the definition \Vhatdef, effectively considering the OPEs of $\hat \Vi$ as if they still
contained the plane wave factor, and moving any plane waves out of the OPE bracket.
This means that the superfields $\hat K_i$ contained in $\hat \Vi$
{\it effectively\/} satisfy
\eqn\effope{
[\Pi^m, \hat K_i]_1 = [\p X^m, \hat K_i]_1 = -2\ap ik^m_i \hat K_i\,.
}
This separation of plane waves was done implicitly in the massless tree-level calculations reviewed in
\treereview. For convenience, we define
\eqn\KNijdef{
\KN_{ij} = :e^{ik_i\cdot X}::e^{ik_j\cdot X}:\,,\qquad
\KN_{ijk} = :e^{ik_i\cdot X}::e^{ik_j\cdot X}::e^{ik_k\cdot X}:\,,
}
where $k_i\cdot X$ means $k_i\cdot X(z_i)$.

\newsubsec\tptkin Three-point kinematics and Koba-Nielsen factor

We parameterize the three particles $i,j,k$ in a open-string three-point correlator with arbitrary mass as
\eqn\momphase{
k_i^2 = -{w_i\over\ap},\quad 2\ap k_i\cdot k_j = \ap(k_k^2 - k_i^2 - k_j^2) = -w_k+w_i+w_j.
}
In terms of this parameterization, the three-point open-string Koba-Nielsen factor
$\KN_{w_1w_2w_3}=z_{12}^{2\ap k_1\cdot k_2}z_{13}^{2\ap k_1\cdot k_3}z_{23}^{2\ap k_2\cdot k_3}$
for external states of mass-level $(w_1,w_2,w_3)$ becomes
\eqn\KNs{
\KN_{w_1w_2w_3} = z_{12}^{-w_3+w_1+w_2}z_{13}^{-w_2+w_1+w_3}z_{23}^{-w_1+w_2+w_3}.
}
For example, $\KN_{100} = {z_{12}z_{13}\over z_{23}}$,
$\KN_{110} = z_{12}^2$,
$\KN_{111} = z_{12}z_{13}z_{23}$.

\newsubsec\ampvscorr The pure spinor formalism prescription

In the pure spinor formalism \psf, the prescription to compute the tree-level
three-point color-ordered amplitude
$A_{w_1w_2w_3}(1,2,3)$ is given by
\eqn\treepresc{
A_{w_1w_2w_3}(1,2,3) = \langle \V1(z_1)\V2(z_2)\V3(z_3) \rangle,
}
where the trace over the Chan-Paton factors is not explicitly included and
$(z_1,z_2,z_3)$ can be fixed to three arbitrary positions due to $SL(2,{\Bbb R})$ invariance.
The fields with non-vanishing conformal weight must be integrated out from \treepresc\ using their
OPEs, while the remaining zero modes of $\l$ and $\t$ are integrated out using the $\langle (\l^3
\t^5)\rangle=1$ prescription \psf, which defines the pure spinor cohomology bracket $\langle \cdot
\rangle$ \PSspace.

The corresponding three-point color-dressed amplitude is defined as
\eqn\coldressed{
A_{w_1w_2w_3} = A_{w_1w_2w_3}(1,2,3){\rm tr}\big(T^{a_1}T^{a_2}T^{a_3}\big)
+ A_{w_1w_3w_2}(1,3,2){\rm tr}\big(T^{a_1}T^{a_3}T^{a_2}\big)
}
where each color-ordered amplitude is associated with a Chan-Paton trace $\tr\big(T^{a_1}T^{a_2}T^{a_3}\big)$.
The gauge group is defined by the commutation relation $[T^a,T^b] = if^{abc}T^c$ where $f^{abc}$
is the structure constant, and the generators are normalized by $\tr\big(T^aT^b\big)=\half \d^{ab}$.
Moreover $d^{abc}=\tr\big(T^aT^bT^c\big)+\tr\big(T^aT^cT^b\big)$ is the symmetrized trace\foot{See
\refs{\Vcolor,\eulerian} for the general decomposition of traces into structure constants and
symmetrized traces.}.


From now on we will drop the color-order specification $(1,2,3)$ and denote the color-ordered
amplitude $A_{w_1w_2w_3}(1,2,3)$
the three-point {\it amplitude} $A_{w_1w_2w_3}$. The distinction with the color-dressed amplitude
\coldressed\ will arise from the context.


\newnewsec\allAsec Manifest M\"obius invariance of massive $3$-point amplitudes

In this section we will show that tree-level three-point massive amplitudes in the pure spinor
formalism can be rewritten in a manifestly
M\"obius-invariant manner.
In order to do this, we will use BRST-cohomology manipulations in pure spinor superspace together with
OPE bracket identities for non-free fields.

\newsubsec\besec BRST exact OPE brackets

Explicit calculations using the vertex operators \Vdefs\ and their worldsheet 
derivatives \delVs\ yields
\eqnn\exids
$$\eqalignno{
[V^{(0)}_1,\p V^{(0)}_2]_1 &= -\ap(k_1+k_2)^2 [V_1^{(0)}, V_2^{(0)}]_0 &\exids\cr
[V_1^{(1)},\p V_2^{(0)}]_2 &= -\ap(k_1+k_2)^2 [V_1^{(1)}, V_2^{(0)}]_1\cr
[V_1^{(1)},\p V_2^{(1)}]_3 &= -\ap(k_1+k_2)^2 [V_1^{(1)}, V_2^{(1)}]_2,
}$$
where $k_i^2 = -w_i/\ap$ for the supermultiplet represented by $V_i^{(w_i)}$. Note that the OPEs
between the plane waves has not been evaluated in the above, so each instance of a superfield
contains its associated plane wave.

To verify these formulas, one plugs in
the explicit forms of the vertices \Vdefs\ and their worldsheet derivatives \delVs\
and evaluate all OPEs in long but straightforward calculations, while paying attention not to
evaluate
the OPEs between the plane waves. Note that the first identity in \exids\ for massless vertices was already
used in \towardsFT, under the same special treatment considerations for plane waves.

Even though the explicit form of the vertices is not known in the pure spinor formalism\foot{The BRST cohomology
in the pure spinor formalism, however,
contains all massive string states in its spectrum \refs{\psspec,\psspecR}. For a
DDF-like construction of vertex operators of arbitrary mass level in the pure spinor formalism
using $SO(8)$ superfields, see \refs{\DDFrenannI,\DDFrenannII}.}
for $w_1>1$, the natural generalization of \exids\
for arbitrary
conformal weights is valid provided one assumption holds: the operator $\p X^m$ arising from the
expansion of a plane wave has non-trivial OPEs only with other plane waves. Together with the
special treatment of plane waves under the OPE bracket formalism \effope, this means that
\eqn\assumption{
[\p X^m, \hat\Vi]_n = -2\ap ik_i^m\hat\Vi \d_{n,1}\,.
}
The generalization of \exids\ can be stated as follows:

\proclaim Proposition.
The OPE bracket
between unintegrated vertices of arbitrary mass levels $\Vi$
and their worldsheet derivatives $\p\Vj$
satisfies
\eqn\bsolv{
[\V{i},\p\V{j}]_{w_i+w_j+1} = -\ap(k_i+k_j)^2 [\Vi,\Vj]_{w_i+w_j}\,.
}\par

\noindent{\it Proof}. The separation of the OPE brackets of ordinary fields and plane waves
using \delV\
\eqn\delVa{
\p\V{j}=\p\hat\V{j}:e^{ik_j\cdot X}: + ik^m_j \hat\V{j} :\p X^m e^{ik_j\cdot X}:,
}
yields
\eqn\lproot{
[\V{i},\p\V{j}]_n =[\hat \V{i},\p\hat\V{j}]_n \KN_{ij}
+ik^m_j[\hat\V{i},\hat\V{j}]_n :e^{ik_i\cdot X}::\p X^m e^{ik_j\cdot X}:
}
But equation \Atwelve\ implies
\eqn\derid{
[\hat \V{i},\p\hat\V{j}]_n =\p [\hat\Vi,\hat\Vj]_{n} +(n-1) [\hat\Vi,\hat\Vj]_{n-1}\,,
}
and the plane-wave OPE gives
\eqn\pwope{
:e^{ik_i\cdot X}::\p X^m e^{ik_j\cdot X}:= :e^{ik_i\cdot X}\p X^m e^{ik_j\cdot X}:
+ 2\ap i k_i^m \KN_{ij}/z_{ij}\,.
}
Therefore \lproot\ becomes
\eqnn\lproo
$$\eqalignno{
[\V{i},\p\V{j}]_n &= \Big(\p [\hat\Vi,\hat\Vj]_{n} +(n-1)
[\hat\Vi,\hat\Vj]_{n-1}\Big)\KN_{ij}&\lproo\cr
&\quad{}+ik^{m}_{j}[\hat\Vi, \hat\Vj]_n :e^{ik_i\cdot X}\p X^m e^{ik_j\cdot X}:\cr
&\quad{}-2\ap k_i\cdot k_j[\hat\Vi, \hat\Vj]_{n-1}\KN_{ij}\,,
}$$
where we reduced the order of the OPE bracket from $n$
to $n{-}1$ in the third line because of the simple pole in the OPE of $\p X^m$ with the plane
wave \pwope.

Setting $n=w_i+w_j+1$ implies that $[\hat\Vi,\hat\Vj]_n=0$ due to \vanB, and yields
\eqnn\bexac
$$\eqalignno{
[\V{i},\p\V{j}]_{w_i+w_j+1} &= (w_i+w_j-2\ap k_i\cdot k_j)[\hat\Vi,\hat\Vj]_{w_i+w_j}\KN_{ij}
&\bexac\cr
&= -\ap(k_i+k_j)^2
[\Vi,\Vj]_{w_i+w_j}\,,
}$$
where $-\ap(k_i+k_j)^2 = (w_i+w_j-2\ap k_i\cdot k_j)$ using \momphase, and we absorbed the
factor $\KN_{ij}$ into the OPE bracket on the RHS for clarity.\qed

Since $\Vi$ is BRST closed and $\p V_j^{(w_j)}$ is BRST exact,
the identity \bsolv\ can be rewritten as
\eqn\bsolvBGt{
[Q,[\Vi,U_j^{(w_j+1)}]_{w_i+w_j+1}]_1 = \ap(k_i+k_j)^2[\Vi,\Vj]_{w_i+w_j}\,,
}
where we used the derivation property \deriv.
This means that $[\Vi,\Vj]_{w_i+w_j}$ is BRST exact if $(k_i+k_j)^2\neq0$,
\eqn\bsolvBG{
[\Vi,\Vj]_{w_i+w_j} =
{1\over \ap(k_i+k_j)^2}[Q,[\Vi,U^{(w_j+1)}_j]_{w_i+w_j+1}]_1\,.
}
The result \bsolvBG\ is the generalization to massive states of an
observation in the massless sector: that the OPE bracket $[\Vz1,\Vz2]_0$ (usually written as
$\Vz1\Vz2$)
is {\it not\/} BRST exact in the massless $3$-point amplitude because $(k_1+k_2)^2=k_3^2=0$ \nptMethod.
More precisely:
\proclaim Lemma. If $k^2_3={-}{w_3\over\ap}\neq0$ then
\eqn\Bvanish{
\langle[[\V1,\V2]_n,\V3]_{w_1+w_2+w_3-n}\rangle = 0,\quad n\ge w_1+w_2\,.
}
\noindent{\it Proof.} When $n=w_1+w_2$ the result follows from \bexacs\ because $\V3$ is BRST
closed and
$[\V1,\V2]_{w_1+w_2}$ is BRST exact when $k^2_3\neq0$ due to \bsolvBG.
When $n>w_1+w_2$ the claim follows from \vanB.\qed

In the appendix~\BGapp, we will use the result \bsolvBG\ to define and obtain massive
Berends-Giele currents of multiplicity two, generalizing the massless definition in \nptMethod.

\newsubsec\moinvsec Manifest M\"obius invariance using level-one vertices

We will first explicitly compute the amplitudes $A_{100}, A_{110}$ and $A_{111}$ using the known
form of the vertices $\Vi$ for $w_i=0,1$ \Vdefs. Using BRST manipulations together with identities obeyed
by the OPE brackets, we will show that these amplitudes are manifestly independent on the vertex
positions at the pure spinor superspace level, i.e. without the need for component expansions.
Then, as a warm-up for a generalization, 
we will consider the calculation of $A_{200}$ and we will show that it requires a new kind of identity among
numerators in pure spinor superspace in order to exhibit manifest M\"obius invariance.
We will then tackle
the general case $A_{w_1w_2w_3}$ of
supermultiplets of arbitrary mass levels, and derive the new numerator identities in the form of a
{\it recurrence relation} (for other types of recurrence relations in the bosonic string, see
\leerev).

\newsubsubsec\Azzoampsec The $A_{100}$ amplitude

The tree-level prescription of the $A_{100}$ amplitude is given by \psf
\eqn\Auzz{
A_{100} = \langle V_1^{(1)}(z_1) V_2^{(0)}(z_2) V_3^{(0)}(z_3)\rangle\,.
}
Separating the Koba-Nielsen factor and extracting the $z_1$ dependence
associated with the conformal weight-one vertex we get
\eqn\Auzzt{
A_{100} =
\Bigl({1\over z_{12}}\langle [[\hat V_1^{(1)}\hat V_2^{(0)}]_1 \hat V_3^{(0)}]_0
-{1\over z_{13}}\langle [\hat V_2^{(0)}[\hat V_1^{(1)}\hat V_3^{(0)}]_1]_0\rangle\Bigr) \KN_{100},
}
where the Koba-Nielsen factor $\KN_{100}={z_{12}z_{13}\over z_{23}}$ is given by \KNs.
Since $(k_2+k_3)^2\neq0$ we know from \Bvanish\ that
$\langle [\hat V_1^{(1)},[\hat V_2^{(0)},\hat V_3^{(0)}]_0]_1\rangle = 0$, therefore
\eqn\prima{
\langle [[\hat\Vo1, \hat\Vz2]_1, \hat\Vz3]_0\rangle -
\langle [\hat\Vz2,[\hat\Vo1, \hat\Vz3]_1]_0\rangle =
\langle[\hat\Vo1, [\hat\Vz2,\hat\Vz3]_0]_1\rangle = 0
}
where we used the derivation
property of the simple pole OPE bracket \farril.
Therefore, we obtain
\eqn\Auzzid{
\langle [\hat\Vz2,[\hat\Vo1,\hat\Vz3]_1]_0\rangle = \langle [[\hat\Vo1, \hat\Vz2]_1, \hat\Vz3]_0\rangle.
}
Plugging \Auzzid\ into the amplitude \Auzz\ leads to
\eqn\Azzoind{
A_{100} = \langle [[\hat\Vo1,\hat\Vz2]_1,\hat\Vz3]_0\rangle\;
\Bigl({1\over z_{12}}-{1\over z_{13}}\Bigr){\rm KN}_{100} =
\langle [[\hat\Vo1,\hat\Vz2]_1,\hat\Vz3]_0\rangle\,.
}
So the amplitude $A_{100}$ is manifestly independent on the vertex positions.

\newsubsubsec\Aoozampsec The $A_{110}$ amplitude

When particles one and two are massive of the first-level and particle three is massless,
the pure spinor tree-level prescription \psf\ yields
\eqn\aooz{
A_{110} = \langle \Vo1(z_1)\Vo2(z_2)\Vz3(z_3)\rangle\,,
}
where the positions $z_1, z_2$ and $z_3$ may be fixed to arbitrary values.
After splitting the Koba-Nielsen factor and extracting the worldsheet positions
associated to conformal weight one (first
$z_1$ followed by $z_2$) we get
\eqnn\Aoozamp
$$\eqalignno{
A_{110} &= \langle [[\hat\Vo1, \hat\Vo2]_2, \hat\Vz3]_0\rangle {\KN_{110}\over z_{12}^2}
+ \langle [[\hat\Vo1, \hat\Vo2]_1, \hat\Vz3]_1\rangle {\KN_{110}\over z_{12}z_{23}} &\Aoozamp\cr
&- \langle [\hat\Vo2, [\hat\Vo1, \hat\Vz3]_1]_1\rangle {\KN_{110}\over z_{13}z_{23}}
}$$
Since particles one and two are massive, $(k_1+k_3)^2\neq0$ and $(k_2+k_3)^2\neq0$ and
therefore $[\hat\Vo1, \hat\Vz3]_1$ and
$[\hat\Vo2, \hat\Vz3]_1$ are BRST exact by \bsolvBG. Thus, by \bexacs
\eqn\twoexs{
\langle [\hat\Vo2, [\hat\Vo1, \hat\Vz3]_1]_1\rangle = 0, \qquad
\langle[\hat\Vo1, [\hat\Vo2, \hat\Vz3]_1]_1 \rangle= 0,
}
which kills the third term in \Aoozamp.
In addition, it follows from the derivation property of the
OPE bracket $[\hat\Vo1, .]_1$ that
\eqn\twobid{
\langle[[\hat\Vo1, \hat\Vo2]_1, \hat\Vz3]_1\rangle = \langle[\hat\Vo1,[\hat\Vo2,
\hat\Vz3]_1]_1\rangle
+ \langle[\hat\Vo2,[\hat\Vo1, \hat\Vz3]_1]_1\rangle= 0\,,
}
which kills the second term in \Aoozamp\ and leads to
\eqn\Aoozampfin{
A_{110} = \langle [[\hat\Vo1, \hat\Vo2]_2, \hat\Vz3]_0\rangle {\KN_{110}\over z_{12}^2} =
\langle [[\hat\Vo1, \hat\Vo2]_2,\hat\Vz3]_0\rangle.
}
Therefore $A_{110}$ is manifestly independent on the vertex positions.

\newsubsubsec\Aoooampsec The $A_{111}$ amplitude

The $A_{111}$ tree-level amplitude is given by
\eqnn\Auuutmp
$$\eqalignno{
A_{111} &= \langle \Vo1(z_1)\Vo2(z_2)\Vo3(z_3)\rangle\cr
&=\Big(
 {\langle [[\hat\Vo1, \hat\Vo2]_2, \hat\Vo3]_1\rangle\over z_{12}^2 z_{23}}
-{\langle [\hat\Vo2, [\hat\Vo1, \hat\Vo3]_2]_1\rangle\over z_{13}^2 z_{23}} &\Auuutmp\cr
&\quad{}+{\langle [[\hat\Vo1, \hat\Vo2]_1, \hat\Vo3]_2\rangle\over z_{12} z_{23}^2}
-{\langle [\hat\Vo2, [\hat\Vo1, \hat\Vo3]_1]_2\rangle\over z_{13} z_{23}^2}
\Big){\rm KN}_{111}
}$$
where $\KN_{111}=z_{12}z_{13}z_{23}$ and we used OPEs to extract the dependence on $z_1$ and $z_2$.

Since $(k_1+k_2)^2\neq0$ and
$(k_1+k_3)^2\neq0$ the OPE brackets $[\hat\Vo1, \hat\Vo2]_2$ and
$[\hat\Vo1, \hat\Vo3]_2$ are BRST exact by \bsolvBG. Thus, using \bexacs\ leads to
\eqn\bexs{
\langle [[\hat\Vo1, \hat\Vo2]_2, \hat\Vo3]_1\rangle = \langle [\hat\Vo2, [\hat\Vo1, \hat\Vo3]_2]_1\rangle = 0,
}
as confirmed by explicit component expansions.
Therefore,
the amplitude \Auuutmp\ becomes
\eqnn\Aoootmp
$$\eqalignno{
A_{111} &=\Big(
{\langle [[\hat\Vo1, \hat\Vo2]_1, \hat\Vo3]_2\rangle\over z_{12} z_{23}^2}
-{\langle [\hat\Vo2, [\hat\Vo1, \hat\Vo3]_1]_2\rangle\over z_{13} z_{23}^2}
\Big)\KN_{111}. &\Aoootmp
}$$
Due to \bsolvBG\ and \bexacs\ we have $\langle [\hat\Vo1, [\hat\Vo2, \hat\Vo3]_2]_1\rangle=0$ because
$(k_2+k_3)^2\neq0$.
The derivation property \deriv\ of $[\hat\Vo1,\cdot]_1$ then leads to
\eqn\gradVo{
0 =\langle[\hat\Vo1, [\hat\Vo2, \hat\Vo3]_2]_1\rangle =
\langle[[\hat\Vo1, \hat\Vo2]_1, \hat\Vo3]_2\rangle
- \langle[\hat\Vo2,[\hat\Vo1, \hat\Vo3]_1]_2\rangle
}
which implies $\langle[\hat\Vo2,[\hat\Vo1, \hat\Vo3]_1]_2\rangle =\langle[[\hat\Vo1, \hat\Vo2]_1, \hat\Vo3]_2\rangle$.
Thus, plugging these numerators into the amplitude \Aoootmp\ and using
\eqn\pfid{
\Big({1\over z_{12} z_{23}^2}-{1\over z_{13} z_{23}^2}\Big) = {1\over z_{12}z_{13}z_{23}}
}
leads to
\eqn\Auuufin{
A_{111} =
\langle [[\hat\Vo1, \hat\Vo2]_1, \hat\Vo3]_2\rangle
{{\rm KN_{111}}\over z_{12}z_{13} z_{23}}
=\langle [[\hat\Vo1, \hat\Vo2]_1, \hat\Vo3]_2\rangle\,.
}
Therefore the amplitude
$A_{111}$ is manifestly independent on the vertex positions.

\newsubsubsec\genAwsec Summary of level-one amplitudes

We demonstrated above that the three-point massive
amplitudes at tree level involving the known vertices $V^{(w)}$ with $w=0,1$ in
the pure spinor formalism are manifestly independent on the
vertex positions. They are given by
\eqnn\mAmps
$$\eqalignno{
A_{100} &= \langle[[\hat\Vo1, \hat\Vz2]_1, \hat\Vz3]_0 \rangle,&\mAmps\cr
A_{110} &= \langle[[\hat V_1^{(1)}, \hat V_2^{(1)}]_2, \hat V_3^{(0)}]_0 \rangle,\cr
A_{111} &= \langle[[\hat V_1^{(1)}, \hat V_2^{(1)}]_1, \hat V_3^{(1)}]_2 \rangle.
}$$
These calculations were confirmed by explicit component expansions, see appendix~\appZ.
In the subsequent, these results will be generalized to arbitrary mass levels $A_{w_1w_2w_3}$.
But before doing that, let us analyze the amplitude $A_{200}$ to see that additional
relations among the numerators are required to obtain all the cancellations necessary
for manifest M\"obius invariance.

\paragraph{The $A_{200}$ amplitude and the need for additional relations} If the mass level
(conformal weights)
of the vertices are $(2,0,0)$, the tree-level $A_{200}$ amplitude is given by
\eqnn\Atwozz
$$\eqalignno{
A_{200} &= \langle \Vt1(z_1)\Vz2(z_2)\Vz3(z_3)\rangle\cr
&=\Big(
 {\langle [[\hat\Vt1, \hat\Vz2]_2, \hat\Vz3]_0\rangle\over z_{12}^2}
-{\langle [\hat\Vz2, [\hat\Vt1, \hat\Vz3]_2]_0\rangle\over z_{13}^2} &\Atwozz\cr
&\quad{}+{\langle [[\hat\Vt1, \hat\Vz2]_1, \hat\Vz3]_1\rangle\over z_{12} z_{23}}
-{\langle [\hat\Vz2, [\hat\Vt1, \hat\Vz3]_1]_1\rangle\over z_{13} z_{23}}
\Big){\rm KN}_{200}
}$$
where $\KN_{200}=z_{12}^2z_{13}^2/z_{23}^2$ and we used OPEs to extract the functional dependence
on $z_1$, followed by other OPEs if possible.

The derivation property \deriv\ implies
\eqn\ntnq{
[\hat\Vt1,[\hat\Vz2,\hat\Vz3]_1]_1 =
[[\hat\Vt1,\hat\Vz2]_1,\hat\Vz3]_1
- [\hat\Vz2,[\hat\Vt1,\hat\Vz3]_1]_1\,.
}
However, the lhs of \ntnq\ vanishes as $[\hat\Vz2,\hat\Vz3]_1=0$ due to \vanB. Therefore,
\eqn\primrel{
[\hat\Vz2,[\hat\Vt1,\hat\Vz3]_1]_1 = [[\hat\Vt1,\hat\Vz2]_1,\hat\Vz3]_1\,.
}
Plugging \primrel\ into \Atwozz\ leads to
\eqn\AtwozzA{
A_{200}=\Big(
 {\langle [[\hat\Vt1, \hat\Vz2]_2, \hat\Vz3]_0\rangle\over z_{12}^2}
-{\langle [\hat\Vz2, [\hat\Vt1, \hat\Vz3]_2]_0\rangle\over z_{13}^2}
+{\langle [[\hat\Vt1, \hat\Vz2]_1, \hat\Vz3]_1\rangle\over z_{12} z_{13}}
\Big){\rm KN}_{200}
}
where we used ${1\over z_{12}z_{23}}-{1\over z_{13}z_{23}} = {1\over z_{12}z_{13}}$ to obtain the last term.

Note that $[\hat\Vz2,\hat\Vz3]_0$ is BRST exact since $(k_2+k_3)^2\neq0$. In addition, as
$\hat\Vt1$ is BRST closed\foot{Recall that for practical
purposes, one can
consider $\hat \Vi$ as $\Vi$ in OPE calculations provided care is taken to exclude the
Koba-Nielsen factor in the end.}, this
implies
the vanishing of $\langle [\hat\Vt1,[\hat\Vz2,\hat\Vz3]_0]_2\rangle$. And now another relation
among the numerators arises from the application of \Asix:
\eqn\newr{
0\sim[\hat\Vt1,[\hat\Vz2,\hat\Vz3]_0]_2 = [[\hat\Vt1,\hat\Vz2]_2,\hat\Vz3]_0
-[\hat\Vz2,[\hat\Vt1,\hat\Vz3]_2]_0 + [[\hat\Vt1,\hat\Vz2]_1,\hat\Vz3]_1
}
and therefore,
\eqn\newrt{
\langle[[\hat\Vt1,\hat\Vz2]_1,\hat\Vz3]_1\rangle =
\langle[\hat\Vz2,[\hat\Vt1,\hat\Vz3]_2]_0\rangle
-\langle[[\hat\Vt1,\hat\Vz2]_2,\hat\Vz3]_0\rangle\,.
}
Plugging \newrt\ into \AtwozzA\ leads to
\eqnn\AtwozzB
$$\eqalignno{
A_{200} &=
 \langle [[\hat\Vt1, \hat\Vz2]_2, \hat\Vz3]_0\rangle\Bigl({1\over z_{12}^2}-{1\over
 z_{12}z_{13}}\Bigr)\KN_{200}&\AtwozzB\cr
&-\langle [\hat\Vz2, [\hat\Vt1, \hat\Vz3]_2]_0\rangle\Bigl({1\over z_{13}^2}-{1\over
z_{12}z_{13}}\Bigr)\KN_{200}\,.
}$$
One can clearly see that \AtwozzB\ is not manifestly M\"obius invariant. Unlike the
amplitudes with level-1 vertices, the {\it naive} relations among numerators that are obtained from BRST
exact superfields and the identities obeyed by the OPE bracket are not sufficient to obtain
manifest independence on the vertex positions. There must be additional relations to show this.
In fact, {\it if\/} the numerators in \AtwozzB\ were related by
\eqn\magrel{
\langle [\hat\Vz2, [\hat\Vt1, \hat\Vz3]_2]_0\rangle \qeq
-\langle [[\hat\Vt1, \hat\Vz2]_2, \hat\Vz3]_0\rangle\,,
}
then the amplitude \AtwozzB\ would be
\eqnn\Atwb
\eqnn\AtwbA
$$\eqalignno{
A_{200} &\qeq \langle [[\hat\Vt1, \hat\Vz2]_2, \hat\Vz3]_0\rangle\Bigl(
{1\over z_{12}^2}+{1\over z_{13}^2}-{2\over z_{12}z_{13}}
\Bigr)\KN_{200}&\Atwb\cr
&=
\langle [[\hat\Vt1, \hat\Vz2]_2, \hat\Vz3]_0\rangle.&\AtwbA
}$$
A result that is manifestly M\"obius invariant.
The natural questions are then:
Is the relation \magrel\ true? If so, how to prove it?

To answer these questions we will derive recurrence relations among
$3$-point massive numerators in \firstp\ and \secp\ and analyze their consequences.
As it turns out, the identity \magrel\ is true, see \twobra\ below for the proof. This means that the
$A_{200}$ amplitude is indeed given by \AtwbA, a functional constant on the worldsheet and therefore
manifestly M\"obius invariant.

In summary, this example shows that {\it additional relations} among
$3$-point massive numerators are needed
to obtain manifestly M\"obius-invariant amplitudes of arbitrary mass levels.
As we will see momentarily, a new recurrence
relation among numerators will give rise to precisely the needed relations.

\newsubsec\genAw The general massive $3$-point amplitude

To compute the massive $3$-point amplitude with arbitrary massive states, we
use OPEs to extract the behavior in $z_1$ followed by $z_2$ to get,
\eqnn\AgenF
$$\eqalignno{
A_{w_{1}w_{2}w_{3}}&=\sum^{w_{1}+w_{2}}_{i=1}
\langle[[\hat\V1, \hat\V2]_{i},\hat\V3]_{w-i}\rangle\, z^{-i}_{12}z^{-w+i}_{23}\,\KN_{w_1w_2w_3}
&\AgenF\cr
&-\sum^{w_{1}+w_{3}}_{i=1}
\langle[\hat\V2, [\hat\V1,\hat\V3]_{i}]_{w-i}\rangle\, z^{-i}_{13}z^{-w+i}_{23}
\,\KN_{w_1w_2w_3}
}$$
where the Koba-Nielsen factor is given in \KNs\ and $w=w_1+w_2+w_3$.

\proclaim Lemma. For $w=w_1+w_2+w_3$, we have
\eqn\trlemma{
\langle [[\hat\V1,\hat\V3]_n,\hat\V2]_{w-n}\rangle =
(-1)^{w-n}\langle [\hat\V2,[\hat\V1,\hat\V3]_n]_{w-n}\rangle, \quad \forall n
}\par

\noindent{\it Proof.} Using \Aten\ we obtain
\eqnn\ptrlem
$$\eqalignno{
[[\hat\V1,\hat\V3]_n,\hat\V2]_{w-n} &= (-1)^{w-n}[\hat\V2,[\hat\V1,\hat\V3]_n]_{w-n} &\ptrlem\cr
&+ (-1)^{w-n}\sum_{i=1}^\infty(-1)^i{1\over i!}\p^i[\hat\V2,[\hat\V1,\hat\V3]_n]_{w-n+i}
}$$
but $\p^i [\hat\V2,[\hat\V1,\hat\V3]_n]_{w-n+i} = 0$ for $i\ge1$ and $\forall n$ due to \vanB.\qed

From now on it will be convenient to define the shorthands specific for the
massive $3$-point amplitude:
\eqnn\Bndef
$$\eqalignno{
P^{12}_n &= \langle [[\hat\V1,\hat\V2]_n,\hat\V3]_{w-n}\rangle\,,\quad n\ge0&\Bndef\cr
P^{13}_n &= \langle [\hat\V2,[\hat\V1,\hat\V3]_n]_{w-n}\rangle\,,\quad n\ge0.
}$$
Note that relabeling $2\leftrightarrow3$ on $P^{12}_n$ leads to
\eqnn\Prela
$$\eqalignno{
P^{12}_n\Big|_{2\leftrightarrow3} &= \langle [[\hat\V1,\hat\V3]_n,\hat\V2]_{w-n}\rangle&\Prela\cr
&=(-1)^{w-n}\langle [\hat\V2,[\hat\V1,\hat\V3]_n]_{w-n}\rangle=(-1)^{w-n}P^{13}_n\,,
}$$
where we used \trlemma\ in the second line.

\newsubsec\recursec Recurrence relations of massive three-point numerators

We will now obtain
a recurrence relation labelled by the pole order $n$ of the three-point numerators $P^{12}_n$ and $P^{13}_n$
from \Bndef. The derivation crucially depends on the BRST cohomology properties of pure spinor superspace
and its pure spinor bracket $\langle\cdot\rangle$.

\proclaim Proposition. The $3$-point pure spinor superspace
OPE brackets \Bndef\ obey the following recurrence relations
\eqnn\firstp
\eqnn\secp
$$\eqalignno{
(2w_1-n)P^{12}_n &= (n-1-w_1-w_2+w_3) P^{12}_{n-1}, \quad n=1,2,3, \ldots &\firstp\cr
(n-2w_1)P^{13}_n &= (n-1-w_1+w_2-w_3) P^{13}_{n-1}\,. &\secp\cr
}$$
%

\noindent{\it Proof}. We will prove the recurrence \firstp\ for $P^{12}_n$, as the recurrence for
$P^{13}_n$ can be obtained by
relabeling.
Setting $i=1$ and $j=2$ and
acting with the operator $[\,\cdot\,,\V3]_{w-n+1}$
on both sides
of \lproo\ with $w=w_1+w_2+w_3$,
we get
\eqnn\init
$$\eqalignno{
0& \sim [\p [\hat\V1,\hat\V2]_{n},\hat\V3]_{w-n+1}\KN_{123}&\init\cr
&+(n-1) [[\hat\V1,\hat\V2]_{n-1},\hat\V3]_{w-n+1}\KN_{123}\cr
&+ik^{m}_{2}[[\hat\V1,\hat\V2]_n],\hat\V3]_{w-n+1}:e^{ik_1\cdot X}\p X^m e^{ik_2\cdot X}::e^{ik_3\cdot X}:\cr
&-2\ap k_{1}\cdot k_{2}[[\hat\V1,\hat\V2]_{n-1},\hat\V3]_{w-n+1}\KN_{123}
}$$
where $0\sim$ on the first line arises because the lhs of \lproo\ is BRST exact, and we made sure
to split the OPE bracket to separate the plane wave factors into the factors of $\KN_{123}$
defined in \KNijdef.
The equation \Aeleven, the plane-wave OPE and
the kinematic relation \momphase\
imply
\eqnn\derida
$$\eqalignno{
[\p [\hat\V1,\hat\V2]_n,\hat\V3]_{w-n+1}&=(n-w)[[\hat\V1,\hat\V2]_n,\hat\V3]_{w-n}&\derida\cr
:e^{ik_1\cdot X}\p X^m e^{ik_2\cdot X}::e^{ik_3\cdot X}: &=
- 2\ap i k_3^m \KN_{123}/z_{23}\cr
-2\ap k_1\cdot k_2 &= w_3-w_1-w_2\,.
}$$
Using these, equation \init\ becomes
\eqnn\tmpf
$$\eqalignno{
0&\sim (n-w_1-w_2-w_3)[[\hat\V1,\hat\V2]_n,\hat\V3]_{w-n} &\tmpf\cr
&+(n-1-w_1-w_2+w_3) [[\hat\V1,\hat\V2]_{n-1},\hat\V3]_{w-n+1}\cr
&+2\ap k_2\cdot k_3[[\hat\V1, \hat\V2]_n,\hat\V3]_{w-n}
}$$
where we reduced the order of the outer OPE bracket in the third line from $w-n+1$ to $w-n$ due to 
the simple pole in \derida.
From \tmpf\ and $2\ap k_2\cdot k_3 = - w_1 +w_2+w_3$ one gets
\eqnn\tmpff
$$\eqalignno{
0&\sim (n-2w_1)[[\hat\V1,\hat\V2]_n,\hat\V3]_{w-n} &\tmpff\cr
&+(n-1-w_1-w_2+w_3) [[\hat\V1,\hat\V2]_{n-1},\hat\V3]_{w-n+1}\,.
}$$
Finally, we transform the BRST exact statement \tmpff\ into an equality by
taking the pure spinor cohomology bracket on both sides of \tmpff\ to obtain,
\eqn\recf{
(2w_1-n)\langle[[\hat\V1,\hat\V2]_n,\hat\V3]_{w-n}\rangle
=(n-1-w_1-w_2+w_3)\langle[[\hat\V1,\hat\V2]_{n-1},\hat\V3]_{w-n+1}\rangle
}
or equivalently, using the definition \Bndef,
\eqn\recff{
(2w_1-n)P^{12}_n =(n-1-w_1-w_2+w_3)P^{12}_{n-1}\,,
}
Equation \recff\ is the recurrence relation \firstp\ we wanted to show.
Furthermore, relabeling $2\leftrightarrow3$ in \recff\ 
and using \Prela\ leads to the recurrence \secp,
\eqn\recs{
(n-2w_1)P^{13}_n =
(n-1-w_1+w_2-w_3)P^{13}_{n-1}\,,
}
finishing the proof. \qed

\paragraph{Classifying the recurrence relations}
Let us denote the recurrence \firstp\ the {\it Recurrence $12$} and \secp\ the {\it Recurrence $13$}.
We will see momentarily that solving the recurrences $12$ and $13$ for general conformal
weights $w_1$, $w_2$ and $w_3$ implies a natural organization of the massive amplitudes
according to the weight distribution among the $w_i$s.

\newsubsubsec\maxsec Maximal non-vanishing pole order

The massive three-point amplitudes are written in terms of the OPE brackets in \Bndef.
It is easy to see that when $n$ is high enough, the corresponding $P^{12}_n$ (or $P^{13}_n$)
is zero.
For example, $P^{12}_{n} = 0$, for $n\ge w_1+w_2$ due to \vanB\ and \Bvanish. Now we would like
to ask the question: what is the
maximal value of $n$ such that $P^{12}_n\neq0$? The answer is given by:

\proclaim Proposition. The recurrence relation \firstp\ together with the vanishing
conditions \Bvanish\
implies that the maximal pole order $n$ for which $P^{12}_n\neq0$
($P^{13}_n\neq0$)
is
\eqn\nmaximal{
n=w_1+w_2-w_3\quad (n=w_1-w_2+w_3).
}

\noindent{\it Proof.} We will prove the claim for $P^{12}_n$ as the case $P^{13}_n$ is similar and
can be obtained by relabeling.
The recurrence relation \firstp\ can be written as
\eqn\Brec{
L_n P^{12}_n = R_n P^{12}_{n-1},\quad n=1,2,3, \ldots
}
where $P^{12}_n$ is the shorthand \Bndef\ and
\eqn\LRdefs{
L_n = (2w_1-n),\qquad R_n = (n-1-w+2w_3)\,.
}
Note that both $L_n$ and $R_n$ can be zero for certain $n$; this leads to a peculiar
behavior of the recurrence \firstp\ at the vanishing points:
\eqnn\mLmRcond
$$\eqalignno{
L_{m_L} = 0 \quad\Longrightarrow\quad P^{12}_n =0,\quad &n\le m_L-1 &\mLmRcond\cr
R_{m_R} = 0 \quad\Longrightarrow\quad P^{12}_n =0,\quad &n\ge m_R\,,
}$$
where
\eqn\mLmR{
m_L = 2w_1\,,\quad m_R = 1+w-2w_3\,.
}
The three logical cases to consider are $m_L>m_R$, $m_L=m_R$ or $m_L<m_R$  (they yield the three
cases in \casesdef\ upon substituting for $w_i$s).
\smallskip
\item{$\bullet$} Case 1: $m_L > m_R$, or equivalently $w_1-w_2+w_3>1$.
The conditions \mLmRcond\ imply that $P^{12}_n=0$ for all $m_R\le n\le m_L -1$.
We claim that $P^{12}_n=0$ when $n$ is above $m_L-1$. To see this, note that if $m_L-1<w_1+w_2$ then
$w_1-w_2<1$ which implies that $w_3>0$ in order for $w_1-w_2+w_3>1$ to be satisfied.
Since $w_3>0$, we have
$P^{12}_{w_1+w_2} = 0$ due to \Bvanish\ or $P^{12}_{n>w_1+w_2} = 0$ due to \vanB.
Therefore, using the recurrence relation \firstp, we conclude
that all values of $n$ above $m_L-1$ lead to $P^{12}_n=0$.
Alternatively, if $m_L-1\ge w_1 + w_2$
then we get $P^{12}_n=0$ due to \vanB. Therefore, in general we have
\eqn\caseOnonI{
P^{12}_n\neq0,\quad n=1, \ldots,w_1+w_2-w_3\,,
}
where the end of the range is given by $m_R-1=w_1+w_2-w_3$, which is the maximal pole order in
\nmaximal.

\item{$\bullet$} Case 2: $m_L = m_R$, or equivalently $w_1-w_2+w_3=1$.
In other words, $m_L=m_R$ has the solution $n_0=2w_1=1+w_1+w_2-w_3$ and
the recurrence relation \Brec\ breaks down at $n_0$; $0\cdot P^{12}_{n_0}=0\cdot P^{12}_{n_0-1}$.
Therefore the maximal $n$
for which $P^{12}_n\neq0$ is either above $n_0$ or it is $n=n_0-1=w_1+w_2-w_3$. But it is easy to see
that the highest value of $n$ that is above $n_0$ is $n_A=w$, for which $P^{12}_{n_A}=0$ due to
\vanB. The recurrence relation \firstp\
then guarantees that $P^{12}_n=0$ for $n>n_0$.
Therefore, in general we have
\eqn\caseOnonII{
P^{12}_n\neq0,\quad n=1, \ldots,w_1+w_2-w_3\,,
}
where the end of the range is $n_0-1=w_1+w_2-w_3$, which is the maximal pole order in \nmaximal.

\item{$\bullet$} Case 3: $m_L < m_R$, or equivalently $w_1-w_2+w_3\le 0$.
We have $P^{12}_n=0$ for $n\ge m_R$ and $n\le m_L-1$, where
$m_L=2w_1$ and $m_R=1+w-2w_3$.
Therefore
\eqn\ranget{
P^{12}_n\neq0, \quad 2w_1\le n\le w_1+w_2-w_3,
}
for which the maximal value of $n$ such that $P^{12}_{w_1+w_2-w_3}\neq0$ is $n=w_1+w_2-w_3$.

\noindent In summary, the recurrence relation \firstp\ and the vanishing condition \Bvanish\ lead to the
conclusion that the maximal order $n$ for a non-vanishing $P^{12}_n$ ($P^{13}_n$) is
$n=w_1+w_2-w_3$ ($w_1-w_2+w_3$), where the maximal $n$ for $P{13}_n$ was obtained by relabeling.\qed

It will be convenient to define
\eqnn\basisf
$$\eqalignno{
P^{12}_{\rm max}&\equiv\langle[[\hat\V1,\hat\V2]_{w_1+w_2-w_3},\hat\V3]_{2w_3}\rangle,
\quad w_1+w_2-w_3\ge1\,,&\basisf\cr
P^{13}_{\rm max}&\equiv\langle[\hat\V2,[\hat\V1,\hat\V3]_{w_1-w_2+w_3}]_{2w_2}\rangle,
\quad w_1-w_2+w_3\ge1\,.
}$$

\paragraph{Types of $w$ distribution}
Note that the analysis above of the recurrence relation for $P^{12}_n$ naturally led to
three distinct cases of conformal weight distribution (similarly for $P^{13}_n$). These cases fall
into two types of inequalities, called {\it types of $w$ distribution\/}. More precisely,
\eqnn\casesdef
\settabs 16 \columns
\+& && Recurrence $12$ &&&&&&& Recurrence $13$ &&&\cr
\+& type $1$ && $w_1-w_2+w_3\ge1$ &&&&& type $1'$ && $w_1+w_2-w_3\ge1$ &&&\cr \hfill\casesdef\hfilneg
\vskip-12pt
\+& type $2$ && $w_1-w_2+w_3\le0$ &&&&& type $2'$ && $w_1+w_2-w_3\le0$ &&&\cr
\smallskip
\noindent From now on, we will see that these types of $w$ distribution naturally induce
the splitting of the correlators, Koba-Nielsen factor, solutions of the recurrence relations etc into
different expressions depending on the choice of the weight distribution $\tau=1,1',2,2'$ for the two
different recurrences
$12$ or $13$ specified in \casesdef.

\newsubsubsec\rrsolsec Closed-form solution of the recurrence relations

For each type $\tau=1,2$ of $w$ distribution \casesdef, one can solve the recurrence
relations \firstp\ and \secp\ in closed form:
\eqnn\solo
\eqnn\solt
$$\eqalignno{
\langle[[\hat\V1, \hat\V2]_{i},\hat\V3]_{w-i}\rangle &= C^{12,(\tau)}_i\;
\langle[[\hat\V1, \hat\V2]_{w_{1}+w_{2}-w_{3}},\hat\V3]_{2w_{3}}\rangle\,,
&\solo\cr
\langle[\hat\V2, [\hat\V1,\hat\V3]_{i}]_{w-i}\rangle &= C^{13,(\tau)}_i\;
\langle[\hat\V2, [\hat\V1,\hat\V3]_{w_{1}+w_{3}-w_{2}}]_{2w_{2}}\rangle\,&\solt
}$$
where  $C^{12,(\tau)}_{i}$ and $C^{13,(\tau)}_{i}$ are given by
\eqnn\Conetwo
\eqnn\Conethree
$$\eqalignno{
C^{12,(\tau)}_i &= \prod^{w_{1}+w_{2}-w_{3}}_{j=i+1}{2w_{1}-j\over j-1-w+2w_{3}}\,,\quad &\Conetwo\cr
C^{13,(\tau)}_i &= \prod^{w_{1}-w_{2}+w_{3}}_{j=i+1} {j-2w_{1}\over j-1-w+2w_{2}}\,. &\Conethree
}$$
The range of the index $i$ on the lhs of \solo\ and \solt\ depends on the type of $w$ distribution \casesdef;
this is because the solution
\solo\ only makes sense if the lhs is non-vanishing.
With some patience, one can rewrite the above products in a closed form for each type individually:
\eqnn\ConetwotypeOne
\eqnn\ConetwotypeTwo
\eqnn\ConethreetypeOne
\eqnn\ConethreetypeTwo
$$\eqalignno{
C^{12,(1)}_i &= {\displaystyle (-1)^{w+i}{2w_1-i-1\choose w-2w_{2}-1}},
\quad i=1,\ldots,w_1+w_2-w_3 - 1, &\ConetwotypeOne\cr
C^{12,(2)}_i &=
{\displaystyle {-w_1+w_{2}-w_{3}\choose i-2w_{1}}},
\quad i=2w_1,\ldots,w_1+w_2-w_3 - 1, &\ConetwotypeTwo\cr
%
C^{13,(1')}_i &= {\displaystyle {2w_1-i-1\choose w-2w_{3}-1}},\quad
i=1,\ldots,w_1-w_2+w_3-1, &\ConethreetypeOne \cr
C^{13,(2')}_i &={\displaystyle (-1)^{w+i}{-w_1-w_{2}+w_{3}\choose i-2w_{1}}},\quad
i=2w_1,\ldots, w_1-w_2+w_3-1, &\ConethreetypeTwo
}$$
In shorthand form using the definitions \Bndef\ and \basisf, the closed-form solutions \solo\ and
\solt\ are written as
\eqn\recsols{
P^{12}_i = C_i^{12,(\tau)}P^{12}_{\rm max},\qquad
P^{13}_i = C_i^{13,(\tau)}P^{13}_{\rm max}\,,\quad i\in R^{(\tau)}
}
where $R^{(\tau)}$ is the set of integers encoding the different ranges according to the type of
$w$ distribution.

Furthermore, the pure spinor BRST cohomology together with the derivation property \deriv\
imply that the two brackets on the RHS of \solo\ and \solt\ are
proportional:

\proclaim Proposition. If $w_1\ge1$ and
$w_1-w_2+w_3\ge1$, $w_1+w_2-w_3\ge1$ ($w$ distribution \casesdef\ is of type $(1,1')$), then
the two maximal pole orders \basisf\ are related as
\eqn\twobra{
\langle[\hat\V2,[\hat\V1,\hat\V3]_{w_1-w_2+w_3}]_{2w_2}\rangle = -
(-1)^{w}\langle[[\hat\V1,\hat\V2]_{w_1+w_2-w_3},\hat\V3]_{2w_3}\rangle\,.
}
or equivalently, $P^{13}_{\rm max} = (-1)^{w+1}P^{12}_{\rm max}$, where $w=w_1+w_2+w_3$.

\noindent{\it Proof.} Consider the OPE bracket derivation identity \deriv,
\eqnn\suder
$$\eqalignno{
[\hat\V1,[\hat\V2,\hat\V3]_{w_1+w_2+w_3-1}]_1 &= [[\hat\V1,\hat\V2]_1,\hat\V3]_{w_1+w_2+w_3-1}
&\suder\cr
&- [\hat\V2,[\hat\V1,\hat\V3]_1]_{w_1+w_2+w_3-1}
}$$
Since $w_1\ge1$ we have $w_1+w_2+w_3-1\ge w_2+w_3$. Using \vanB\ and \Bvanish\ this means that
$[\hat\V1,[\hat\V2,\hat\V3]_{w_1+w_2+w_3-1}]_1$ either vanishes on the spot (if $w_1>1$) or
it is BRST exact (if $w_1=1$).
Therefore, the cohomology of the pure spinor brackets implies
\eqn\exId{
\langle[\hat\V2,[\hat\V1,\hat\V3]_1]_{w_1+w_2+w_3-1}\rangle =
\langle[[\hat\V1,\hat\V2]_1,\hat\V3]_{w_1+w_2+w_3-1}\rangle,
}
or, in other words, $P^{13}_1=P^{12}_1$.
From the solution \solt\ we get
\eqn\solPmaxs{
C^{13,(1')}_1 P^{13}_{\rm max} = P^{13}_1
= P^{12}_1
= C^{12,(1)}_1 P^{12}_{\rm max}
}
therefore, from \ConethreetypeOne\ and \ConetwotypeOne\
and $(w-2w_2-1)+(w-2w_3-1)=2w_1-2$
we obtain
$$
C^{12,(1)}_1 =(-1)^{w+1} {2w_1-2\choose w-2w_{2}-1} =
(-1)^{w+1} {2w_1-2\choose w-2w_{3}-1} =
(-1)^{w+1} C^{13,(1')}_1
$$
that is,
$C_1^{12,(1)}=(-1)^{w+1}C_1^{13,(1')}$ and therefore
\eqn\Pmaxs{
P^{13}_{\rm max} = (-1)^{w+1}P^{12}_{\rm max},\quad{w_1+w_2-w_3\ge1\atop w_1-w_2+w_3\ge1}
}
finishing the
proof.\qed

\newsubsec\genMoesec Manifest M\"obius invariance of massive $3$-point amplitudes

Suppose we want to compute the general massive three-point amplitude at tree-level with the pure
spinor formalism. From the possible OPEs alone, extracting $z_1$ followed by $z_2$, we get
(to avoid the massless amplitude we assume that $w_1\ge1$)
\eqnn\Agen
$$\eqalignno{
A_{w_{1}w_{2}w_{3}}&=\sum^{w_{1}+w_{2}-w_{3}}_{i=1}
\langle[[\hat\V1, \hat\V2]_{i},\hat\V3]_{w-i}\rangle\, z^{-i}_{12}z^{-w+i}_{23}\,\KN_{w_1w_2w_3} &\Agen\cr
&-\sum^{w_{1}-w_{2}+w_{3}}_{i=1}
\langle[\hat\V2, [\hat\V1,\hat\V3]_{i}]_{w-i}\rangle\, z^{-i}_{13}z^{-w+i}_{23}
\,\KN_{w_{1}w_{2}w_{3}}
}$$
where the Koba-Nielsen factor is given in \KNs, and we already used the result \nmaximal\ for
the maximal pole order to restrict the upper limit of the sums. We will refer to the two
terms in \Agen\ as the $12$ and the $13$ channels.

However, having the general correlator is not enough for a complete picture; we know from the
analysis of the recurrence relations that the different values of $w_1$, $w_2$, $w_3$ gives rise
to totally different behavior of the amplitude. More precisely, the analysis of the
Recurrence $12$ for the $12$ channel implied two different $w$ distributions, denoted
$\tau=1,2$ and similarly for the Recurrence $13$ with $\tau=1',2'$
listed in \casesdef. For example, if in the analysis of the Recurrence $12$
we consider the type $2$, $w_1-w_2+w_3\le0$, it means that there is no $13$ channel in the
general amplitude \Agen, as this case invalidates the upper limit of the sum on the second line
of \Agen.

In summary, for each kind of recurrence ($12$ or $13$) there are 2 possibilities \casesdef\ for the $w$
distribution, for a total of
four possibilities $(\tau,\tau')$ when computing $A_{w_1w_2w_3}$.
\eqnn\fpos
\eqnn\fposTw
\eqnn\fposTh
\eqnn\fposF
$$\eqalignno{
(1,1'):\quad & w_1-w_2+w_3\ge1,\quad w_1+w_2-w_3\ge1 &\fpos\cr
(2,1'):\quad &w_1-w_2+w_3\le0,\quad w_1+w_2-w_3\ge1 &\fposTw\cr
(1,2'):\quad &w_1-w_2+w_3\ge1,\quad w_1+w_2-w_3\le0 &\fposTh\cr
(2,2'):\quad &w_1-w_2+w_3\le0,\quad w_1+w_2-w_3\le0 &\fposF
}$$
The range of $i$ in the sums on \Agen\ will change accordingly.
\smallskip

\proclaim Proposition. The general massive three-point amplitude $A_{w_1w_2w_3}$ with
$w_1\ge1$ is manifestly independent on the vertex positions and is given by
\eqn\AgenProp{
A_{w_1w_2w_3}= \cases{
\langle[[\hat\V1,\hat\V2]_{w_{1}+w_{2}-w_{3}},\hat\V3]_{2w_{3}}\rangle & $w\in\{(1,1'),(2,1')\}$ \cr
(-1)^{w+1}\langle [\hat\V2,[\hat\V1,\hat\V3]_{w_1-w_2+w_3}]_{2w_2}\rangle
& $w\in\{(1,2')\}$
}
}
where the different weight distributions are given by \fpos,\fposTw\ and \fposTh.\par

\noindent{\it Proof.}
We will consider the three\foot{The fourth distribution \fposF\
is invalid for massive amplitudes as they imply $w_1\le0$.} types of $w$
distributions \fpos-\fposTh\ in turn: $(1,1')$, $(2,1')$, $(1,2')$, and we will show that they all lead to
the conclusion \AgenProp.

\item{$\bullet$} $w_1-w_2+w_3\ge1$ and $w_1+w_2-w_3\ge1$, a $w$ distribution of type $(1,1')$. There are no
restrictions in both sums of \Agen, and we get
\eqnn\AgenA
$$\eqalignno{
A_{w_{1}w_{2}w_{3}}&=
\sum^{w_{1}+w_{2}-w_{3}}_{i=1}
\langle[[\hat\V1, \hat\V2]_{i},\hat\V3]_{w-i}\rangle\;
z^{-i}_{12}z^{-w+i}_{23}\,\KN_{w_{1}w_{2}w_{3}}\qquad{} &\AgenA\cr
&-\sum^{w_{1}-w_{2}+w_{3}}_{i=1}
\langle[\hat\V2, [\hat\V1,\hat\V3]_{i}]_{w-i}\rangle\, z^{-i}_{13}z^{-w+i}_{23}
\,\KN_{w_1w_2w_3}\cr
%
}$$
\item{$\bullet$} $w_1-w_2+w_3\le0$ and $w_1+w_2-w_3\ge1$, a $w$ distribution of type $(2,1')$.
There is
no valid range of summation in the second line of \Agen\ because $w_1-w_2+w_3\le0$ and we get,
\eqn\Agenctwo{
A_{w_{1}w_{2}w_{3}}=\sum^{w_{1}+w_{2}-w_{3}}_{i=2w_{1}}
\langle[[\hat\V1, \hat\V2]_{i},\hat\V3]_{w-i}\rangle\; z^{-i}_{12}z^{-w+i}_{23}
\KN_{w_1w_2w_3}\,,
}
where the sum starts at $i=2w_1$, as derived in \ranget.
\item{$\bullet$} $w_1-w_2+w_3\ge1$ and $w_1+w_2-w_3\le0$, a $w$ distribution of type $(1,2')$.
There is no valid range for the sum in the first line of \Agen\
$w_1+w_2-w_3\le0$ and we get,
\eqn\Agencthree{
A_{w_1w_2w_3}=-\sum^{w_{1}-w_{2}+w_{3}}_{i=2w_{1}}
\langle[\hat\V2, [\hat\V1,\hat\V3]_i]_{w-i}\rangle\; z^{-i}_{13}z^{-w+i}_{23}
\KN_{w_1w_2w_3}\,.
}
\item{$\bullet$} $w_1-w_2+w_3\le0$ and $w_1+w_2-w_3\le0$, a $w$ distribution of type $(2,2')$.
There is no valid massive solution to both these inequalities as
they imply $2w_1\le0$.

\noindent Now, for the correlator in each of these types, \AgenA, \Agenctwo\ and \Agencthree,  there are different solutions for the rewriting of
the $12$ and $13$ channels in terms
of the maximal poles \Bndef.

For a $w$ distribution of type $(1,1')$:
$w_1-w_2+w_3\ge1$, and $w_1+w_2-w_3\ge1$. We use the solutions \ConetwotypeOne\ and \ConethreetypeOne\ of the recurrence relations
\eqn\PsolsOO{
P_i^{12} = C_i^{12,(1)}P^{12}_{\rm max},\quad
P_i^{13} = C_i^{13,(1')}P^{13}_{\rm max}
}
in the type $(1,1')$ amplitude \AgenA\ to get
\eqnn\AgenB
$$\eqalignno{
A_{w_{1}w_{2}w_{3}}^{(1,1')}&=
\Big(P^{12}_{\rm max}\;H^{12,(1)}
-P^{13}_{\rm max}\; H^{13,(1')}\Big)\KN_{w_{1}w_{2}w_{3}}\cr
}$$
where we defined
\eqnn\Hdefs
$$\eqalignno{
H^{12,(1)}&= \sum^{w-2w_{3}}_{i=1}C^{12,(1)}_i z^{-i}_{12}z^{-w+i}_{23} =
\sum^{w-2w_{3}}_{i=1}(-1)^{w+i}{2w_1-i-1\choose w-2w_{2}-1}z^{-i}_{12}z^{-w+i}_{23}\,,&\Hdefs\cr
H^{13,(1')}&=\sum^{w-2w_{2}}_{i=1}C^{13,(1')}_i z^{-i}_{13}z^{-w+i}_{23} =
\sum^{w-2w_{2}}_{i=1}{2w_1-i-1\choose w-2w_{3}-1}z^{-i}_{13}z^{-w+i}_{23}\,.
}$$
Finally, since for type $(1,1')$ we have $w_1-w_2+w_3\ge1$ and $w_1+w_2-w_3\ge1$, both $P^{12}_{\rm max}$
and $P^{13}_{\rm max}$ exist and they are related by
the identity \twobra\ as $P_{\rm max}^{13}=(-1)^{w+1}P^{12}_{\rm max}$. Therefore,
\eqn\AgenFin{
A_{w_1w_2w_3}^{(1,1')} =
P^{12}_{\rm max}\,
\big(H^{12,(1)}+(-1)^wH^{13,(1')}\big)\KN_{w_1w_2w_3} = P^{12}_{\rm max}
}
where we used $\big(H^{12,(1)}+(-1)^wH^{13,(1')}\big)\KN_{w_1w_2w_3}=1$ as will be shown in
the appendix~\appComb.

For a $w$ distribution of type $(2,1')$, we use the solution \ConetwotypeTwo
\eqn\PsolTO{
P^{12}_i = C_i^{12,(2)}P^{12}_{\rm max}
}
in the amplitude \Agenctwo\ to get
\eqn\AgenctwoA{
A_{w_{1}w_{2}w_{3}}^{(2,1')}  = P^{12}_{\rm max}H^{12,(2)}\KN_{w_1w_2w_3} = P^{12}_{\rm max}
}
%
%
%
where we defined
\eqn\Honetwop{
H^{12,(2)}=\sum^{w-2w_{3}}_{i=2w_{1}}C^{12,(2)}_i z^{-i}_{12}z^{-w+i}_{23}=
\sum^{w-2w_{3}}_{i=2w_{1}}z^{-i}_{12}z^{-w+i}_{23}{-w_1+w_{2}-w_{3}\choose i-2w_{1}}
}
and used that $H^{12,(2)}\KN_{w_1w_2w_3}=1$, whose proof is in the appendix~\appComb.

For a $w$ distribution of type $(1,2')$, we use the solution \ConethreetypeTwo
\eqn\PsolTT{
P^{13}_i = C_i^{13,(2')}P^{13}_{\rm max}
}
in the amplitude \Agencthree\ to get
\eqn\AOT{
A_{w_1w_2w_3}^{(1,2')}  = -P^{13}_{\rm max}H^{13,(2')}\KN_{w_1w_2w_3} = (-1)^{w+1}P^{13}_{\rm max}
}
where we defined
\eqn\Honethreep{
H^{13,(2')}=\sum^{w_1-w_{2}+w_3}_{i=2w_{1}}C^{13,(2')}_i z^{-i}_{13}z^{-w+i}_{23}=
\sum^{w_1-w_2+w_3}_{i=2w_{1}}z^{-i}_{13}z^{-w+i}_{23}(-1)^{w+i}{-w_1-w_{2}+w_{3}\choose i-2w_{1}}
}
and used that $H^{13,(2')}\KN_{w_1w_2w_3}=(-1)^w$, whose proof is in the appendix~\appComb.

As already mentioned above,
in the appendix~\appComb\ we will show that for $w_1\ge1$, the following equations are true
\eqnn\Hids
$$\eqalignno{
\big(H^{12,(1)}+(-1)^wH^{13,(1')}\big)\KN_{w_1w_2w_3}&=1\,,\quad
&\Hids\cr
H^{12,(2)}\KN_{w_1w_2w_3}&=1\cr
H^{13,(2')}\KN_{w_1w_2w_3}&=(-1)^w\,.
}$$
In summary, the general massive amplitude for all different $w$ distributions with $w_1\ge1$
is universally given by
\eqn\AgF{
A_{w_1w_2w_3}= \cases{
\langle[[\hat\V1,\hat\V2]_{w_{1}+w_{2}-w_{3}},\hat\V3]_{2w_{3}}\rangle & $w\in\{(1,1'),(2,1')\}$ \cr
(-1)^{w+1}\langle [\hat\V2,[\hat\V1,\hat\V3]_{w_1-w_2+w_3}]_{2w_2}\rangle
& $w\in\{(1,1'),(1,2')\}$
}
}
and it is {\it manifestly} independent on the vertex positions.\qed

Note that if the weight distribution is of type $(1,1')$ given in \fpos, then both lines of
\AgF\ agree with each other, as proven in \twobra.

\newsubsec\exsec Examples

We will now showcase some example amplitudes for various different $w$ distributions, highlighting
the non-trivial cancellations of the vertex positions on the worldsheet.

\paragraph{The $A_{211}$ amplitude} We have $(w_1,w_2,w_3)=(2,1,1)$, the total weight
is $w=4$ and
$w_1-w_2+w_3 = 2$,
$w_1+w_2-w_3 = 2$,
which means that this amplitude is of type $(1,1')$, see \fpos. 
The combination of worldsheet singularities in \AgenFin\ and
the Koba-Nielsen factor \KNs\
are given by
\eqn\kntoo{
H^{12,(1)}+H^{13,(1')} =
{1\over z_{12}^2 z_{23}^2}
- {2\over z_{12} z_{23}^3}
+ {1\over z_{13}^2 z_{23}^2}
+ {2\over z_{13} z_{23}^3},\quad
\KN_{211} = z_{12}^2z_{13}^2\,.
}
One can check that $\big(H^{12,(1)}+H^{13,(1')}\big)\KN_{211} = 1$, as predicted by our method. To
see this
non-trivial cancellation, one uses the partial-fraction techniques explained in \tdef.
The
amplitude \AgenProp\ becomes
\eqn\exdoo{
A_{211} = \langle[[\hat V_1^{(2)},\hat V_2^{(1)}]_2, \hat V_3^{(1)}]_2\rangle\,.
}

\paragraph{The $A_{251}$ amplitude}
We have $(w_1,w_2,w_3)=(2,5,1)$, the total weight is $w=8$ and
$w_1-w_2+w_3 = -2$,
$w_1+w_2-w_3 = 6$,
which means that this amplitude is of type $(2,1')$, see \fposTw.
The combination \Honetwop\ of worldsheet singularities and
the Koba-Nielsen factor \KNs\ are given by
\eqn\KNddz{
H^{12,(2)} = {1\over z_{12}^6 z_{23}^2}
       + {2\over z_{12}^5 z_{23}^3}
       + {1\over z_{12}^4 z_{23}^4},\quad
\KN_{251} = {z_{12}^6 z_{23}^4\over z_{13}^2}\,,
}
from which it follows that $H^{12,(2)}\KN_{251}=1$. The amplitude \AgenProp\ becomes
\eqn\exddz{
A_{251} = \langle[[\hat V_1^{(2)},\hat V_2^{(5)}]_6, \hat V_3^{(1)}]_2\rangle\,.
}

\paragraph{The $A_{214}$ amplitude}
We have $(w_1,w_2,w_3)=(2,1,4)$, the total weight is $w=7$ and
$w_1-w_2+w_3 = 5$,
$w_1+w_2-w_3 = -1$,
which means that this amplitude is of type $(1,2')$, see \fposTh.
The combination \Honethreep\ of worldsheet singularities and
the Koba-Nielsen factor \KNs\ are given by
\eqn\kntof{
H^{13,(2')} = {1\over z_{13}^5z_{23}^2}
       - {1\over z_{13}^4z_{23}^3}\,,\quad
\KN_{214} =  {z_{13}^5 z_{23}^3\over z_{12}}\,.
}
One can check that $H^{13,(2')}\KN_{214}=(-1)^7=-1$ and the amplitude \AgenProp\ becomes
\eqn\amptotp{
A_{214} = \langle [\hat V_2^{(1)},[\hat V_1^{(2)},\hat V_3^{(4)}]_{5}]_{2}\rangle\,.
}

\newsubsec\cdsec The color-dressed amplitude

The color-dressed amplitude \coldressed\ sums over the two inequivalent cyclic permutations of
three strings, namely
\eqn\colordressed{
A_{w_1w_2w_3} = A_{w_1w_2w_3}(1,2,3){\rm tr}\big(T^{a_1}T^{a_2}T^{a_3}\big)
+ A_{w_1w_3w_2}(1,3,2){\rm tr}\big(T^{a_1}T^{a_3}T^{a_2}\big)\,.
}
The color-ordered amplitude in the orderings $(1,2,3)$ and $(1,3,2)$ are given by \AgF\ and its
permutation
\eqnn\AgFt
$$\eqalignno{
A_{w_1w_2w_3}(1,2,3) &= \cases{
\langle[[\hat\V1,\hat\V2]_{w_{1}+w_{2}-w_{3}},\hat\V3]_{2w_{3}}\rangle & $w_1+w_2-w_3\ge1$ \cr
(-1)^{w+1}\langle [\hat\V2,[\hat\V1,\hat\V3]_{w_1-w_2+w_3}]_{2w_2}\rangle & $w_1-w_2+w_3\ge1$
}\cr
A_{w_1w_3w_2}(1,3,2) &= \cases{
\langle[[\hat\V1,\hat\V3]_{w_{1}+w_{3}-w_{2}},\hat\V2]_{2w_{2}}\rangle & $w_1-w_2+w_3\ge1$ \cr
(-1)^{w+1}\langle [\hat\V3,[\hat\V1,\hat\V2]_{w_1+w_2-w_3}]_{2w_3}\rangle & $w_1+w_2-w_3\ge1$
}\cr
&&\AgFt
}$$
Therefore, the computation of the color-dressed amplitude \colordressed\ splits into two cases,
depending on the $w$ distribution: if $w_1+w_2-w_3\ge1$ we have
\eqnn\Aodt
$$\eqalignno{
A_{w_1w_2w_3}(1,2,3)&=\langle[[\hat\V1,\hat\V2]_{w_{1}+w_{2}-w_{3}},\hat\V3]_{2w_{3}}\rangle
&\Aodt\cr
A_{w_1w_3w_2}(1,3,2)&=(-1)^{w+1}\langle [\hat\V3,[\hat\V1,\hat\V2]_{w_1+w_2-w_3}]_{2w_3}\rangle\cr
&=(-1)^{w+1}\langle[[\hat\V1,\hat\V2]_{w_{1}+w_{2}-w_{3}},\hat\V3]_{2w_{3}}\rangle\cr
&=(-1)^{w+1} A_{w_1w_2w_3}(1,2,3)
}$$
where we used \trlemma\ on the third line. The other case if $w_1-w_2+w_3\ge1$ is similar and the
same result follows. This means that we have successfully derived the worldsheet parity factor
$(-1)^{1+w_1+w_2+w_3}$, in accordance with \refs{\schwarzrev,\openstrings}.

Using \Aodt, the color-dressed amplitude \colordressed\ becomes
\eqn\cdtwo{
A_{w_1w_2w_3} = A_{w_1w_2w_3}(1,2,3)\big(\tr(T^1 T^2 T^3 + (-1)^{1+w}\tr(T^1T^3T^2)\big),
}
where $T^1=T^{a_1}$,
and we obtain two results depending on whether the total weight $w$ is even or odd:
\eqn\cdressedfinal{
A_{w_1w_2w_3} = \cases{{i\over 2}f^{123}A_{w_1w_2w_3}(1,2,3), & $w$ even\cr
2d^{123}A_{w_1w_2w_3}(1,2,3), & $w$ odd
}}
where $A_{w_1w_2w_3}(1,2,3)$ is given by \AgFt.

\newnewsec\concsec Conclusions

In this paper, we used the pure spinor tree-level prescription to obtain an expression for the
color-ordered three-point amplitude of arbitrary mass level specified by a triplet of conformal weights $(w_1,w_2,w_3)$,
\eqn\conceq{
A_{w_1w_2w_3}(1,2,3) = \cases{
\langle[[\hat\V1,\hat\V2]_{w_1+w_2-w_3},\hat\V3]_{2w_3}\rangle & $w_1+w_2-w_3\ge1$ \cr
(-1)^{w+1}\langle [\hat\V2,[\hat\V1,\hat\V3]_{w_1-w_2+w_3}]_{2w_2}\rangle
& $w_1-w_2+w_3\ge1$
}
}
where the vertices $\hat\Vi$ have conformal weight $w_i$ at zero momentum, the OPE bracket
$[\cdot,\cdot]_n$ captures the order $n$ singularity of non-free fields, and $\langle\cdot\rangle$
represents the pure spinor cohomology bracket \refs{\psf,\PSspace}.
The salient feature of such a result is its manifest $SL(2,\Bbb R)$ (M\"obius) invariance, being a
constant on the worldsheet with no dependence on the vertex positions $z$.

The derivation of this result relied heavily on the BRST cohomology properties of pure spinor
superspace. We exploited it to obtain recurrence relations for the superfield numerators that
capture the different singularity orders among the vertex operators $\hat\V1$, $\hat\V2$, and
$\hat\V3$ of the three-point amplitude. More precisely, the recurrence relation
\eqn\recODconc{
\langle[[\hat\V1,\hat\V2]_n,\hat\V3]_{w-n}\rangle = {R_n\over L_n}
\langle[[\hat\V1,\hat\V2]_{n-1},\hat\V3]_{w-(n-1)}\rangle
}
where $L_n$ and $R_n$ are defined in \LRdefs,
related poles of order $n$ and $n{-}1$ in the OPE of $\hat\V1$ and $\hat\V2$ (channel $12$); a similar expression
was derived in \secp\ for the other OPE channel $13$.
The solution to \recODconc\ led to a one-dimensional basis for the
channel $12$. Upon plugging this solution into the amplitude (written with all sorts of
$1/(z-w)^n$ poles from OPEs
and $z-w$ dependencies from the Koba-Nielsen factor), we demonstrated that all the $z$ dependence
cancels out, see \Hids (for a non-trivial example cancellation, see \KNddz). Several
technical details had to be overcome, and these are explained in the main text. As a consistency
check of the formula \conceq, we used it to obtain the well-known result that the color-dressed
amplitude of massive states is proportional to either a structure constant or a symmetrized trace
depending on whether the total mass level is even or odd, see \cdressedfinal.

The main achievement of this paper is the M\"obius-invariant expression \conceq.

\bigskip
\noindent{\bf Acknowledgements:}
YT would like to thank Nordita for hosting him during several
stages of the project. YT is supported by the National Natural Science Foundation of China (NSFC) (Grant
NO.\ 124B2094) and the National Key R\&D Program of China (NO.\ 2020YFA0713000).

\appendix{A}{Inverse Koba-Nielsen factor of massive states}
\applab\appComb

\noindent In this appendix we will demonstrate the equations \Hids\ that are fundamental to the proof of
manifest M\"obius invariance of the massive three point amplitudes \AgenProp. The equations \Hids\ give
the explicit form of
the inverse Koba-Nielsen factor corresponding to each of the possible conformal weight distributions
listed in \casesdef.

\paragraph{Trivializing partial fractions}
In order to incorporate the
partial fraction identity, we define
\eqn\tdef{
t={z_{12}\over z_{13}}\quad\Longrightarrow\quad z_{12}=\frac{t}{1-t}z_{23},\quad z_{13}=\frac{1}{1-t}z_{23},
}
Therefore the partial fraction identity is trivially satisfied in this language
\eqn\pfast{
\frac{1}{z_{12}z_{23}}+\frac{1}{z_{23}z_{31}}+\frac{1}{z_{31}z_{12}}=
z^{-2}_{23}\Big(\frac{1-t}{t}+\frac{t^{2}-t}{t}-\frac{(t-1)^{2}}{t}\Big)=0\,.
}
In this notation, the Koba-Nielsen factor \KNs\ and its inverse can be rewritten as
\eqnn\KNast
\eqnn\iKNast
$$\eqalignno{
\KN_{w_{1}w_{2}w_{3}}&=
t^{w_1+w_2-w_3}(1-t)^{-2w_1}z_{23}^{w} &\KNast\cr
\KN^{-1}_{w_{1}w_{2}w_{3}}&=t^{w_3-w_1-w_2}(1-t)^{2w_1}z_{23}^{-w}=
(-1)^w z_{23}^{-w}\sum_{i=0}^{2w_1}{2w_1\choose i}(-t)^{i-w_1-w_2+w_3}&\iKNast
}$$
where we used the binomial theorem to obtain the polynomial form of the
inverse Koba-Nielsen -- an expression which
will be needed below.

Using the variables \tdef,
the definitions \Hdefs\ become
\eqnn\Honetwoast
\eqnn\Honethreeast
$$\eqalignno{
H^{12,(1)}&=(-1)^{w}z^{-w}_{23}\sum^{w-2w_{3}}_{i=1}(1-\frac{1}{t})^{i}{2w_1-i-1\choose w-2w_{2}-1},
&\Honetwoast\cr
H^{13,(1')}&=z^{-w}_{23}\sum^{w-2w_{2}}_{i=1}(1-t)^{i}{2w_1-i-1\choose w-2w_{3}-1},&\Honethreeast
}$$
We are now ready to prove:

\proclaim Proposition. If $w_1\ge1$
then
\eqnn\difc
\eqnn\dift
\eqnn\diff
$$\eqalignno{
\Big((-1)^{w}H^{13,(1')}+H^{12,(1)}\Big)&=\KN_{w_1w_2w_3}^{-1}\,,&\difc\cr
H^{12,(2)}&=\KN_{w_1w_2w_3}^{-1}\,,&\dift\cr
H^{13,(2')}&=(-1)^w \KN_{w_1w_2w_3}^{-1}&\diff
}$$
where $H^{12,(1)}$, $H^{13,(1')}$, $H^{12,(2)}$, $H^{13,(2')}$  are given in \Honetwoast, \Honethreeast,
\Honetwop, and \Honethreep\ and $\KN_{w_1w_2w_3}^{-1}$
in \iKNast. 
\par

\noindent{\it Proof.}
Using the binomial theorem, we get
\eqn\binH{
\Big((-1)^{w}H^{13,(1')}+H^{12,(1)}\Big)=(-1)^{w}z_{23}^{-w} P(t)
}
where $P(t)$ is a polynomial in $t$ given by
\eqn\polyt{
P(t) = \bigg[\sum_{j=1}^{w_1+a}{2w_1-j-1\choose w_1-a-1}
\sum^{j}_{k=0}{j\choose k}(-\frac{1}{t})^{k}
+\sum_{j=1}^{w_1-a}{2w_1-j-1\choose w_1+a-1}\sum^{j}_{k=0}{j\choose k}(-t)^{k}\bigg]
}
and we defined $a=w_2-w_3$. Note that, by hypothesis,
$w_1+a=w_1+w_2-w_3\ge1$ and
$w_1-a=w_1-w_2+w_3\ge1$, therefore the upper range of the sums in \polyt\
are always valid. We will show that the
polynomial $P(t)$ is the same as the polynomial for the inverse Koba-Nielsen factor in \iKNast.
The constant term $t^0$ arises from the $k=0$ term in the
sums
\eqnn\extmp
$$\eqalignno{
\sum_{j=1}^{w_1+a}{2w_1-j-1\choose w_1-a-1}
&=\sum_{j=0}^{w_1+a-1}{2w_1-j-2\choose w_1+a-j-1} =
\sum_{k=0}^{w_1+a-1}{m+k\choose k}\cr
\sum_{j=1}^{w_1-a}{2w_1-j-1\choose w_1+a-1}&=
\sum_{j=0}^{w_1-a-1}{2w_1-j-2\choose w_1-j-a-1}=
\sum_{l=0}^{w_1-a-1}{n+l\choose l} &\extmp
}$$
where we used ${A\choose B}={A\choose A-B}$ and
redefined $m=w_{1}-a-1$, $n=w_{1}+a-1$, $k=n-j$ and $l=m-j$.
Using the identity
$\sum_{k=0}^{r}{m+k\choose k}={m+r+1\choose r}$
yields
\eqn\finalp{
\sum_{j=1}^{w_1+a}{2w_1-j-1\choose w_1-a-1}+\sum_{j=1}^{w_1-a}{2w_1-j-1\choose w_1+a-1}=
{2w_{1}-1 \choose w_{1}+a-1}+{2w_{1}-1 \choose w_{1}+a}={2w_1 \choose w_1+a}
}
where we used ${2w_1-1\choose w_1-a-1}={2w_1-1\choose w_1+a}$ in the second
term and applied ${m\choose n-1}+{m\choose n}={m+1\choose n}$ to arrive at the
last equality. Therefore, the polynomial \polyt\ becomes
\eqn\septmp{
P(t)=
P_1(t)
+ P_2(t)
+ {2w_1\choose w_1+a}
}
where 
\eqnn\Ponedef
\eqnn\Ptwodef
$$\eqalignno{
P_1(t) &= \sum_{j=1}^{w_1+a}\sum_{k=1}^{j}{2w_1-j-1\choose w_1-a-1}{j\choose
k}\(-\frac{1}{t}\)^{k}&\Ponedef\cr
P_2(t)&=\sum_{j=1}^{w_1-a}\sum_{k=1}^{j}{2w_1-j-1\choose w_1+a-1}{j\choose k}(-t)^{k}
&\Ptwodef
}$$
Inverting the order of summation in the sums \Ponedef\ and \Ptwodef\ yields
\eqnn\Pone
\eqnn\Ptwo
$$\eqalignno{
P_1(t)&=
\sum_{k_0=1}^{w_1+a}\sum^{w_1+a-k_{0}}_{i=0}{2w_1-1-i-k_{0}\choose w_1-a-1}
{i+k_{0}\choose k_{0}}\(-\frac{1}{t}\)^{k_{0}} &\Pone\cr
P_2(t) &=
\sum_{k_0=1}^{w_1-a}\sum^{w_1-a-k_{0}}_{i=0}{2w_1-1-i-k_{0}\choose w_1+a-1}{i+k_{0}\choose k_{0}}(-t)^{k_{0}}
&\Ptwo
}$$
The inner sums can be simplified by
redefining $m=w_{1}+a-1$ $n=w_{1}-a-1$, $l=2w_{1}-1$:
$$\eqalignno{
\sum^{w_1+a-k_{0}}_{i=0}{2w_1-1-i-k_{0}\choose w_1-a-1}{i+k_{0}\choose
k_{0}}=&\sum^{l-k_{0}-n}_{i=0}{l-i-k_{0}\choose n}{i+k_{0}\choose k_{0}}=
{2 w_1\choose k_0+w_1-a}
\cr
\sum^{w_1-a-k_{0}}_{i=0}{2w_1-1-i-k_{0}\choose w_1+a-1}{i+k_{0}\choose
k_{0}}=&\sum^{l-k_{0}-m}_{i=0}{l-i-k_{0}\choose m}{i+k_{0}\choose k_{0}}=
{2 w_1\choose k_0+w_1+a}
}$$
where we used \knuthconcrete
\eqn\ntcombid{
\sum^{l-m-n}_{k= 0}{l-m-k\choose n}{k+m\choose m}={l+1\choose m+n+1}\,.
}
Therefore, the non-constant terms \Pone\ and \Ptwo\ can
be rewritten as
\eqnn\Pfins
\eqnn\Ptwofins
$$\eqalignno{
P_1(t) &= \sum_{k_0=1}^{w_1+a}{2w_1\choose a - k_0 + w_1}\(-{1\over t}\)^{k_0}\,
= \sum^{w_1+a-1}_{i=0}{2w_1\choose i}\(-{1\over t}\)^{w_1+a-i}
&\Pfins\cr
P_2(t) &= \sum_{k_0=1}^{w_1-a}{2 w_1\choose k_0+w_1+a}(-t)^{k_0} =
\sum_{i=w_1+a+1}^{2w_1}{2w_1\choose i}(-t)^{i-w_1-a}
&\Ptwofins
}$$
where in $P_1(t)$ we used ${2 w_1\choose k_0+w_1-a}={2 w_1\choose a- k_0+w_1}$
followed by the change of the sum dummy index to $i=w_1+a-k_0$.\foot{This
change of variables modify the bounds of the sum to be $\sum_{i=w_1+a-1}^{0}$,
but we use the convention of summing over variables with an {\it increasing} range.}
We obtain
\eqnn\Ptfim
$$\eqalignno{
P(t) &= P_1(t) + {2w_1\choose w_1+a} + P_2(t)\cr
&= \sum_{i=0}^{w_1+a-1}{2w_1\choose i}(-t)^{i-w_1-a}
+ {2w_1\choose w_1+a}
+ \sum_{i=w_1+a+1}^{2w_1}{2w_1\choose i}(-t)^{i-w_1-a}\cr
&=\sum_{i=0}^{2w_1}{2w_1\choose i}(-t)^{i-w_1-a}\,.&\Ptfim
}$$
Finally, using \Ptfim\ implies that \binH\ becomes
\eqn\binHf{
\Big((-1)^{w}H^{13,(1')}+H^{12,(1)}\Big)=(-1)^{w}z_{23}^{-w}\sum_{i=0}^{2w_1}{2w_1\choose
i}(-t)^{i-w_1-w_2+w_3}\,,
}
where we replaced $a=w_2-w_3$. Comparing \iKNast\ with \binHf\ proves our assertion \difc.

Now we will prove \dift. We have
\eqn\diftt{
\sum^{w-2w_{3}}_{i=2w_{1}}z^{-i}_{12}z^{-w+i}_{23}{-w_1+w_{2}-w_{3}\choose i-2w_{1}}=
z_{23}^{-w}\sum^{w-2w_{3}}_{j=2w_{1}}(-1)^{j}{-w_1+w_{2}-w_{3}\choose j-2w_{1}}\sum^{j}_{k=0}{j\choose k}\(-\frac{1}{t}\)^{k}
}
where we used $z_{12}^{-j}=z_{23}^{-j}\big({1-t\over t}\big)^j=(-1)^j z_{23}^{-j}\big(1-{1\over t}\big)^j$ followed
by the binomial theorem. In the formula above, the terms of order $(-1/t)^{k_0}$ are given by
\eqnn\loid
$$\eqalignno{
\sum^{w-2w_{3}-k_{0}}_{i=2w_{1}-k_{0}}\!\!\!(-1)^{i+k_{0}}{-w_1+w_{2}-w_{3}\choose i+k_{0}-2w_{1}}
{i+k_{0}\choose k_{0}}
&=\sum^{n}_{j=0}(-1)^{j}{n\choose j}{m+j\choose k_{0}} &\loid\cr
&=(-1)^{-w}{2 w_{1} \choose k_{0}+w_{1}-w_{2}+w_{3}}
}$$
where we defined $n=-w_{1}+w_{2}-w_{3}$, $j=i+k_{0}-2w_{1}$, $m=2w_{1}$ and used \knuthconcrete
\eqn\morecomb{
\sum^{N}_{i=0}(-1)^{i}{N \choose i}{M+i\choose k}=(-1)^{N}{M \choose k-N}
}
on the last line of \loid.
Summing all possible powers of $t$, we get:
\eqnn\sumpo
$$\eqalignno{
\sum^{w-2w_{3}}_{j=2w_{1}}(-1)^{j}{-w_1+w_{2}-w_{3}\choose j-2w_{1}}\sum^{j}_{k=0}{j\choose k}(-\frac{1}{t})^{k}
&=(-1)^{-w}\sum^{w_{1}+a}_{k_{0}=0}{2 w_{1} \choose -k_{0}+w_{1}+a}(-t)^{-k_{0}}\cr
&=(-1)^{-w}\sum^{2w_{1}}_{i=0}{2 w_{1} \choose i}(-t)^{-w_{1}-a+i}&\sumpo
}$$
where we used $a=w_2-w_3$ and defined $i=-k_{0}+w_{1}+a$ after using ${A\choose B}={A\choose A-B}$ on the first
line.
Since the hypothesis assumes $w_1-w_2+w_3\le0$, we know that $2w_{1}\leq w_{1}+w_2-w_3=w_1+a$.
The binomial coefficient implies that all $i>2w_{1}$ terms in the sum vanish. Finally we have,
\eqn\finpro{
z^{-w}_{23}\Big(\sum^{w-2w_{3}}_{j=2w_{1}}(-1)^{j}{-w_1+w_{2}-w_{3}\choose j-2w_{1}}
\sum^{j}_{k=0}{j\choose k}(-\frac{1}{t})^{k}\Big)=
(-1)^{-w}z^{-w}_{23}\sum^{2w_{1}}_{i=0}{2 w_{1} \choose i}(-t)^{-w_{1}-w_2+w_3+i}
}
which, after comparing with \iKNast, finishes the proof of \dift.

To prove \diff, we start by expanding the
factor of $z_{13}^{-i} = (1-t)^i z_{23}^{-i}$ in the
expression \Honethreep\ using the binomial theorem
\eqnn\Hotst
$$\eqalignno{
H^{13,(2')}&=
\sum^{w_1-w_2+w_3}_{i=2w_{1}} (-1)^{w+i}{-w_1-w_{2}+w_{3}\choose i-2w_{1}}
z^{-i}_{13}z^{-w+i}_{23} &\Hotst\cr
&=
\sum^{w_1-w_2+w_3}_{i=2w_{1}}(-1)^{w+i}{-w_1-w_{2}+w_{3}\choose i-2w_{1}}
\sum_{k=0}^i{i\choose k} (-t)^k z^{-w}_{23}\cr
&=
(-1)^w\sum_{k_0=0}^{w_1-w_2+w_3}
\sum_{j=2w_1-k_0}^{w_1-w_2+w_3-k_0}(-1)^{j+k_0}
{-w+2w_{3}\choose j+k_0-2w_1}{j+k_0\choose k_0}(-t)^{k_0}z_{23}^{-w}
}$$
Setting $M=2w_1$, $N=-w+2w_3$,$\ell=-2w_1+j+k_0$, the inner sum becomes \knuthconcrete
\eqn\idnt{
\sum_{\ell=0}^N(-1)^\ell{N\choose \ell}{M+\ell\choose k_0}=(-1)^N{M\choose k_0-N}
}
therefore
\eqn\shocmb{
\sum_{j=2w_1-k_0}^{w_1-w_2+w_3-k_0}(-1)^{j+k_0}
{-w+2w_{3}\choose j+k_0-2w_1}{j+k_0\choose k_0}=(-1)^{-w}{2w_1\choose k_0+w-2w_3}\,.
}
Using \shocmb\ in \Hotst\ yields
\eqn\htfin{
H^{13,(2')}=
z_{23}^{-w}\sum_{k_0=0}^{w_1-w_2+w_3}
{2w_1\choose k_0+w-2w_3}
(-t)^{k_0}
=z_{23}^{-w}\sum_{i=w-2w_3}^{2w_1}{2w_1\choose i}(-t)^{i-w+2w_3}
}
where after setting $i=k_0+w-2w_3$ we obtain a sum from $i=w-2w_3$ to
$i=2w_1$. Note however, that $w-2w_3=w_1+w_2-w_3\le0$ by hypothesis, see the $2'$ $w$ distribution
in \fposTh. Therefore the sum really starts from $i=0$ and we get
\eqn\hpfin{
H^{13,(2')}=
z_{23}^{-w}\sum_{i=0}^{2w_1}{2w_1\choose i}(-t)^{i-w+2w_3}=(-1)^w \KN_{w_1w_2w_3}^{-1}\,,
}
which finishes the proof.\qed

\appendix{B}{Superfields in the first-level massive integrated vertex}
\applab\tensupapp

\noindent The integrated vertex \intver\ of the first-level massive states in the pure spinor formalism was
derived in
\intUpaper. The explicit form of the ten superfields that appear in the
definition \intver\
are given by (note that our conventions are different),
\eqnn\tensup
$$\eqalignno{
F_{mn} &= -  9 G_{mn}, &\tensup\cr
F_{m}{}^{\a} &=  -  2 (\g^{k}H^m)^{\a} i \ap,\cr
G_{m\a} &=   3 H^{m}_{\a} \cr
F_{mnp} &=
  3 B_{mnp}
 +  9 G_{mn} k_{p} i \ap
-  9 G_{mp} k_{n} i \ap\cr
K^{\a\b} &=
-  {1 \over 2}\, B_{mnp} (\g^{mnp})^{\a\b} \ap\cr
F^{\a}{}_{\b} &=
- 8\ap C^\a{}_\b\cr
G^{\a}{}_{mn} &=
 {1 \over 6}  (\g^{n}H^m)^{\a} \ap
 + {2 \over 3}  k^{m}(\g^{k}H^n)^{\a} \ap^2
- (m\leftrightarrow n)\cr
H_{\a\b} &=
B_{\a\b}\cr
H_{mn\a} &=
  i \ap\Big[ 2 k^m H^{n}_{\a}
 +  {1 \over 2} (\g^{k m}H^n)_{\a}{}
 - (m\leftrightarrow n)\Big]\cr
G_{mnpq} &=
 {i\ap \over 2}\Big[ B_{npq} k_{m}
 -   B_{mpq} k_{n}
-   B_{pmn} k_{q}
+   B_{qmn} k_{p}\Big] \cr&
+  {3\ap^2 \over 2}\Big[
    G_{nq} k_{m} k_{p}
 -  G_{mq} k_{n} k_{p}
 -  G_{np} k_{m} k_{q}
 +  G_{mp} k_{n} k_{q}  \Big]\,.
}$$
Their equations of motion can be inferred from the equations of motion of the fundamental
superfields given in \eomtheta.

\appendix{C}{Superspace form of massive amplitudes with level-1 states}
\applab\appZ

\noindent In this appendix, we exploit the known forms of the unintegrated vertices of massless and
first-level massive states to compute the amplitudes $A_{100}$, $A_{110}$ and $A_{111}$ using
the general result \AgF. As usual, beware of the usage between $\Vi$ and $\hat\Vi$ in view of
\effope\ and whether the Koba-Nielsen factor has already been canceled or not.

\newsubsubsec\AozzPSSsec The $A_{100}$ amplitude

The relevant OPE bracket  was computed in \UV
\eqnn\VuoVt
$$\eqalignno{
[\Vo1, \Vz2]_1 &= -2\ap(\l H_1^m)\p_m V_2 + 2\ap (C_1\l)^\b D_\b V_2
-{\ap\over2}(\l F_1)_{mn}(\l\g^{mn}A_2) &\VuoVt\cr
& =
2\ap (\l H_1^m)(\l\g^m W_2) + 2\ap Q\Big( (\l H^m_1)A^m_2\Big)
-2\ap Q\Big( (C_1\l)^\a A^2_\a\Big)\cr
}$$
therefore, as the pure spinor cohomology bracket $\langle \cdot\rangle$ annihilates BRST-exact
terms, we get
\eqn\AozzAmp{
A_{100} = \langle [[\hat\Vo1, \hat\Vz2]_1, \hat\Vz3]_0 \rangle = 2\ap \langle (\l H_1^m)(\l\g^m W_2)V_3\rangle.
}
The bosonic component expansion of \AozzAmp, with the normalization $\langle(\l^3\t^5)\rangle=1$,
is \PSS
\eqn\Auzzcomp{
A_{100} = {1\over 320}\ap g_{1mn}\big(f_2^{ma}f_3^{na} + f_2^{na}f_3^{ma}\big)
        + {1\over 320}i b_{1mnp}e_3^m f_2^{np}\,,
}
where $g_{mn}$ is the symmetric traceless $2$-tensor and $b_{mnp}$ is the $3$-form polarization
encoding the first-level massive supermultiplet
\BCpaper\
subject to $k^m g_{mn} = k^m b_{mnp} = 0$, and $f_j^{mn} = k^m_j e^n_j - k^n_j e^m_j$ is
the component SYM field-strength of the vector polarization $e_j^m$.

This amplitude was firstly computed in the pure spinor formalism in \PSthreemass, but the
cancellation of the vertex positions was only accomplished after performing the component expansion.
In \UV\ this cancellation was made manifest already at the superfield level.
See also \massSweden\ for a derivation using field-theory methods with corresponding
interpretation in terms of Berends-Giele \BerendsME\ currents.

\newsubsubsec\AoozPSSsec The $A_{110}$ amplitude

We will evaluate the brackets in
\eqn\wantAooz{
A_{110} = \langle [[\hat\Vo1, \hat\Vo2]_2, \hat\Vz3]_0\rangle.
}
After a long calculation, the double pole in the OPE of two unintegrated vertices of the first massive level is
\eqnn\dpole
$$
\eqalignno{
[\Vo1,\Vo2]_2 & =
\ap\big[  2 (\l B_2 C_1\l)
 -  (\l H^{m}_{1}) (\l H^{m}_{2}) \big] &\dpole\cr
&
+  \ap^2\Big[4 (C_{1}\l)^{\a} \big(D_{\a}(\l H_{2}^{k^{1}})\big) i
 -   (C_{1}\g^{mn}\l)^\a \big(D_{\a}(\l F_{2})_{mn}\big)
 -  4  (C_{1}\l)^{\a} (\g^{k^1}C_{2}\l)_{\a} i\Big]\cr&
 + \ap^2\Big[ 2 \big(D_{\a}(C_{1}\l)^{\b}\big) \big(D_{\b}(C_{2}\l)^{\a}\big)
 +  2 (\l H^{k^{2}}_{1}) (\l H^{k^{1}}_{2}) -  (\l F_{1})_{mn} (\l\g^{mn}H^{k^{1}}_{2}) i \Big]\cr&
 + \ap ^2\Big[ 5(\l F_{1})_{mn} (\l F_{2})_{mn}
 -  {1 \over 8} (\l\g^{mn} F_{1})_{pq} (\l\g^{pq} F_{2})_{mn}\Big]
 - (1\leftrightarrow2)
}$$
After multiplying on the right by $\hat\Vz3$ we obtain the superfields in the
brackets of \wantAooz. However, we can simplify the result
by noting that $\langle [\hat\Vo1, \hat\Vo2]'_2, \hat\Vz3]_0\rangle = \langle [\hat\Vo1, \hat\Vo2]_2, \hat\Vz3]_0\rangle$
where
$[\hat\Vo1, \hat\Vo2]'_2 = [\hat\Vo1, \hat\Vo2]_2 - Q\Lambda_{12}$,
since $\langle [Q\Lambda_{12},\hat\Vz3]_0\rangle = 0$.
Choosing,
\eqnn\Lambdadef
$$\eqalignno{
\Lambda_{12} &= \ap^2\Big[
  4 H^{k^{2}}_{1\; \a} (C_{2}\l)^\a i
+  {4 \over 3}\, H^{m}_{1\; \a} (C_{2}\g^{k^{1} m}\l )^\a i
-  36 (\l H^{m}_{1}) G^{2}_{k^{1}m} i &\Lambdadef\cr&
 +  {1 \over 2}\, B^{1}_{mnp} (\l\g^{npk^1k^2} H_2^m) 
 -  {1 \over 6}\, B^{1}_{mnp} (\l\g^{mnpk^2} H_2^{k^1})\Big]
-(1\leftrightarrow2)
}$$
we get
\eqnn\tptS
$$\eqalignno{
[\Vo1, \Vo2]'_2 & =
\ap^2\Big[
 {5\over2} (\l H^{k^{2}}_{1}) (\l H^{k^{1}}_{2})
 +  3 (\l H^{m}_{1}) (\l\g^{mk^{2}}H^{k^{1}}_{2})
 +  (\l\g^{mk^{2}}H^{n}_{1}) (\l\g^{nk^{1}}H^{m}_{2})\Big]\cr&
 - 2\ap \Big[(\l H^{m}_{1}) (\l H^{m}_{2}) + (\l B_1 C_2\l)\Big]
-(1\leftrightarrow2) &\tptS
}$$
where we used $(k_1\cdot k_2)=1/\ap$ and the equations of motion of the various superfields.
So the amplitude \wantAooz\ can be equivalently written as
$A_{110} = \langle [[\hat\Vo1, \hat\Vo2]'_2, \hat\Vz3]_0\rangle$, leading to
\eqnn\AooAmp
$$\eqalignno{
A_{110} &=
 {5\over2}\ap^2 \langle(\l H^{k^{2}}_{1}) (\l H^{k^{1}}_{2}) V_3\rangle
 +  3\ap^2 \langle (\l H^{m}_{1}) (\l\g^{mk^{2}}H^{k^{1}}_{2}) V_3\rangle &\AooAmp\cr&
 +  \ap^2\langle(\l\g^{mk^{2}}H^{n}_{1}) (\l\g^{nk^{1}}H^{m}_{2})V_3\rangle\cr&
 - 2\ap \langle(\l H^{m}_{1}) (\l H^{m}_{2})V_3\rangle -2\ap\langle (\l B_1 C_2\l)V_3\rangle
-(1\leftrightarrow2)
}$$
To show that the expression \AooAmp\ is BRST closed, in addition to the usual identities
coming from the pure spinor constraint and the Berkovits-Chandia gauge on the superfields,
one also needs the identities\foot{Note that we are using the Schoonschip shorthand notation
popularized by FORM \FORM, where
the name of a vector (e.g. $k^1$, $k^2$) appears in place of its contracted vectorial index. For
example $(\l\g^{nk^{1}}H^{m}_{2})=(\l\g^{np}H^{m}_{2})k^1_p$.}
that follow from the Garnir symmetry $(\l\g^{[s})_\a(\l\g^{mnpqr]}\l)=0$
\eqnn\bido
\eqnn\bidt
$$\eqalignno{
& B^j_{mnp}(\l\g^{nk^1}H_i^q)(\l\g^{mpqk^1k^2}\l) =  B^{j}_{k^{1}mp} (\l
H^{q}_{i}) (\l \g^{mpqk^{1}k^{2}} \l) &\bido\cr
&{} -  {2 \over 3}\, B^{j}_{mnp} (\l H^{k^{1}}_{i}) (\l \g^{mnpk^{1}k^{2}}\l) 
 +  {1 \over 3}\, B^{j}_{mnp} (\l H^{q}_{i}) (\l \g^{mnpqk^{2}} \l) (k^1\cdot k^1)\cr
&{} +  {1 \over 3}\, B^{j}_{mnp} (\l\g^{k^{1}k^{2}}H^{q}_{i}) (\l\g^{mnpqk^{1}} \l)
 -  {1 \over 3}\, B^{j}_{mnp} (\l H^{q}_{i}) (\l \g^{mnpqk^{1}} \l) (k^1 \cdot k^2)
}$$
$$\eqalignno{
& B^{j}_{mnp} (\l\g^{nk^{2}}H^{q}_{i}) (\l \g^{mpqk^{1}k^{2}} \l)  =
B^{j}_{k^{2}mp} (\l H^{q}_{i}) (\l \g^{mpqk^{1}k^{2}} \l) &\bidt\cr&
-  {2 \over 3}\, B^{j}_{mnp} (\l H^{k^{2}}_{i}) (\l \g^{mnpk^{1}k^{2}}\l) 
 +  {1 \over 3}\, B^{j}_{mnp} (\l H^{q}_{i}) (\l \g^{mnpqk^{2}} \l) (k^1\cdot k^2) \cr&
 +  {1 \over 3}\, B^{j}_{mnp} (\l\g^{k^{1}k^{2}}H^{q}_{i}) (\l\g^{mnpqk^{2}} \l) 
-  {1 \over 3}\, B^{j}_{mnp} (\l H^{q}_{i}) (\l \g^{mnpqk^{1}} \l) (k^2 \cdot k^2)
}$$
One can check that \AooAmp\ yields the bosonic component expansion
\eqnn\tptmassA
$$\eqalignno{
A_{110} &=
{9 \over 80}i\ap^2\big[\, g_{1}^{k^{2}k^{2}} g_{2}^{k^{1}e^3}
 -  g_{1}^{k^{2}e^3} g_{2}^{k^{1}k^{1}}
 +  g_{1}^{mk^{2}} g_{2}^{mk^{1}} (k^1 \cdot e^3) \big] &\tptmassA\cr
&+  {9 \over 80}i\ap\big[ g_{1}^{me^3} g_{2}^{mk^{1}}
-   g_{1}^{mk^{2}} g_{2}^{me^3}
-  \half g_{1}^{mn} g_{2}^{mn} (k^1 \cdot e^3)\big]\cr
&+  {9 \over 80}\ap\big[ g_{1}^{mk^{2}} b_{2}^{mk^{1}e^3}
 -  g_{2}^{mk^{1}} b_{1}^{mk^{2}e^3}\big]\cr
&+  {9 \over 160}i\big[ b_{1}^{mne^3} b_{2}^{mnk^{1}}
 -  b_{1}^{mnk^{2}} b_{2}^{mne^3}
-  {1 \over 3} b_{1}^{mnp} b_{2}^{mnp} (k^1 \cdot e^3)\big]
}$$

\newsubsubsec\AoooAmpsec The $A_{111}$ amplitude

After a long and tedious OPE calculation we obtain an expression for
$[[\Vo1 \Vo2]_1 \Vo3]_2$ that contains 222 terms.
We leave the simplification of these terms in pure spinor superspace to a
future work.

It is, however, straightforward to compute its bosonic component
expansion:
$$\eqalignno{
&{320\over 81}\langle[[\hat\Vo1, \hat\Vo2]_1,\hat\Vo3]_2\rangle =
(2\ap)   \Big(
{2 \over 3}\, g_{1}^{k^{2}k^{2}} b_{2}^{mnp}b_{3}^{mnp} \cr&
+  2 g_{1}^{mk^{2}} b_{2}^{mnp} b_{3}^{npk^{1}}
 -  2 g_{1}^{mk^{2}} b_{2}^{npk^{1}} b_{3}^{mnp}
 +  3 g_{1}^{mn} g_{2}^{mp} g_{3}^{np} \cr&
 -  2 g_{1}^{mn} g_{2}^{mp} b_{3}^{npk^{1}} i
 -  2 g_{1}^{mn} g_{3}^{mp} b_{2}^{npk^{1}} i 
 +  {2 \over 3}\, g_{2}^{k^{1}k^{1}} b_{1}^{mnp}b_{3}^{mnp}\cr&
 -  2 g_{2}^{mk^{1}} b_{1}^{mnp} b_{3}^{npk^{1}} 
 -  2 g_{2}^{mk^{1}} b_{1}^{npk^{2}} b_{3}^{mnp} 
 -  2 g_{2}^{mn} g_{3}^{mp} b_{1}^{npk^{2}} i \cr&
  +  {2 \over 3}\, g_{3}^{k^{1}k^{1}} b_{1}^{mnp} b_{2}^{mnp}
  -  2 g_{3}^{mk^{1}} b_{1}^{mnp} b_{2}^{npk^{1}}
  +  2 g_{3}^{mk^{1}} b_{1}^{npk^{2}} b_{2}^{mnp}  \Big)\cr&
+  (2\ap)^2   \Big(  g_{1}^{k^{2}k^{2}} g_{2}^{mn} g_{3}^{mn}
  +  3 g_{1}^{mk^{2}} g_{2}^{mn} g_{3}^{nk^{1}}
  -  3 g_{1}^{mk^{2}} g_{2}^{nk^{1}} g_{3}^{mn}\cr&
 +  2 g_{1}^{mk^{2}} g_{2}^{nk^{1}} b_{3}^{mnk^{1}} i
 -  2 g_{1}^{mk^{2}} g_{3}^{nk^{1}} b_{2}^{mnk^{1}} i
 +  g_{1}^{mn} g_{2}^{k^{1}k^{1}} g_{3}^{mn} \cr&
 -  3 g_{1}^{mn} g_{2}^{mk^{1}} g_{3}^{nk^{1}}
 +  g_{1}^{mn} g_{2}^{mn} g_{3}^{k^{1}k^{1}}
 +  2 g_{2}^{mk^{1}} g_{3}^{nk^{1}} b_{1}^{mnk^{2}} i\Big)\cr&
 +  (2\ap)^3 \Big(
 -  g_{1}^{k^{2}k^{2}} g_{2}^{mk^{1}} g_{3}^{mk^{1}}
 +  g_{1}^{mk^{2}} g_{2}^{k^{1}k^{1}} g_{3}^{mk^{1}}
 -  g_{1}^{mk^{2}} g_{2}^{mk^{1}} g_{3}^{k^{1}k^{1}}\Big)\cr&
 +  2 g_{1}^{mn} b_{2}^{mpq} b_{3}^{npq}
 +  2 g_{2}^{mn} b_{1}^{mpq} b_{3}^{npq}
 +  2 g_{3}^{mn} b_{1}^{mpq} b_{2}^{npq}\cr&
 -  4 b_{1}^{mnk^{2}} b_{2}^{mpq} b_{3}^{npq} i
 -  4 b_{1}^{mnp} b_{2}^{mnq} b_{3}^{pqk^{1}} i
 -  4 b_{1}^{mnp} b_{2}^{mqk^{1}} b_{3}^{npq} i
}$$
This agrees, up to convention changes, with the RNS calculations of
\stiet\foot{Reference \stiet\ restricts the bosonic components to the
polarizations $g_{mn}$ and omits $b_{mnp}$.}.

\appendix{D}{Massive Berends-Giele currents in superspace}
\applab\BGapp

\noindent In this appendix, we use the formula \bsolvBG\ to define massive Berends-Giele currents of
multiplicity two. This is a generalization of how Berends-Giele currents are constructed in the
massless amplitudes, see the review \treereview.

We define
\eqnn\bimplies
$$\eqalignno{
[\Vz1, \Vz2]_0 &= QM^{(0,0)}_{12}, \quad\hbox{if $(k_1+k_2)^2\neq0$} &\bimplies\cr
[\Vo1, \Vz2]_1 &= QM^{(1,0)}_{12}, \quad\hbox{if $(k_1+k_2)^2\neq0$}\cr
[\Vo1, \Vo2]_2 &= QM^{(1,1)}_{12}, \quad\hbox{if $(k_1+k_2)^2\neq0$}.
}$$
More explicitly, from \bsolvBG\ we get
\eqn\BGsm{
M^{(0,0)}_{12}={[\Vz1, U^{(1)}_2]_1\over \ap(k_1+k_2)^2},\quad
M^{(1,0)}_{1 2}={[\Vo1, U^{(1)}_2]_2\over \ap(k_1+k_2)^2},\quad
M^{(1,1)}_{1 2}={[\Vo1, U^{(2)}_2]_3\over \ap(k_1+k_2)^2}
}
The first equation corresponds to the well-known massless Berends-Giele current
reviewed in \treereview.

\newsubsec\bgone Berends-Giele current for one massive and one massless multiplet

According to \BGsm\ and \bimplies, the Berends-Giele current
\eqn\BGom{
M^{(1,0)}_{1 2}={[\Vo1, U^{(1)}_2]_2\over \ap(k_1+k_2)^2}
}
satisfies the BRST variation
\eqn\QMuot{
QM^{(1,0)}_{1 2} = [\Vo1, \Vz2]_1.
}
In principle, one can simply compute the
double pole bracket,
\eqnn\brdou
$$\eqalignno{
[\Vo1,  U^{(1)}_2]_2 & =
   4 (\l H^{k^{2}}_{1}) A_{2}^{k^{1}} \ap ^2 
 -  2 (\l H^{m}_{1}) A_{2}^{m} \ap  
 +  2 (\l\g^{mk^{1}}H^{n}_{1}) F^{2}_{mn} i \ap ^2 &\brdou\cr&
 -  (\l\g^{mn}H^{k^{2}}_{1}) F^{2}_{mn} i \ap ^2 
 +  9 B^{1}_{k^{2}mn} (\l \g^{k^{1}mn}  W^2) \ap ^2 
 +  {1 \over 2}\, B^{1}_{mnp} (\l \g^{k^{1}mnp}  A^2) i \ap  \cr&
 -  B^{1}_{mnp} (\l \g^{k^{1}mnp}  A^2) (k^1 \cdot k^2) i \ap ^2 
 -  2 B^{1}_{mnp} (\l \g^{mnp}  W^2) \ap  \cr&
 -  B^{1}_{mnp} (\l \g^{mnp}  W^2) (k^1 \cdot k^2) \ap ^2 
 +  18 G^{1}_{k^{2}m} (\l \g^{m}  W^2) i \ap ^2
}$$
and be done with it as it satisfies
\eqn\QVU{
Q[\Vo1,  U^{(1)}_2]_2 = \ap(k_1+k_2)^2[\Vo1, \Vz2]_1\,,
}
with $[\Vo1, \Vz2]_1$ given in \VuoVt.
But we can do a bit more by exploiting the cohomology of
pure spinor superspace. First, as can be seen in \VuoVt, there are BRST exact terms in the right-hand
side of \QVU. They can be absorbed by redefining the left-hand side of \QVU\
$$
L^{(1,0)}_{1 2}=[\Vo1,  U^{(1)}_2]_2
- 2\ap^2(k_1+k_2)^2(\l H^m_1)A^m_2
+ 2\ap^2(k_1+k_2)^2(C_1\l)^\a A^2_\a
$$
now satisfies $QL^{(1,0)}_{1 2} = \ap(k_1+k_2)^2\big(2\ap (\l H_1^m)(\l\g^m W_2)\big)$,
corresponding to the BRST-equivalent bracket in \AozzAmp.
Furthermore, subtracting another BRST-exact term by defining
\eqn\VuoVzt{
V^{(1,0)}_{1 2} = L^{(1,0)}_{1 2} - Q\big(4i\ap^2 H_{1\a}^m W^{m\a}_2\big)
}
where $W^{m\a}_2= k_2^m W_2^\a$ is a higher-mass single-particle superfield \hSYM, we obtain a
simple expression,
\eqnn\Vuddef
$$\eqalignno{
V^{(1,0)}_{1 2} &=
  4 (\l H^{k^{2}}_{1}) A_{2}^{k^{1}} \ap ^2
 -  4 (\l H^{m}_{1}) A_{2}^{m} (k^1 \cdot k^2) \ap ^2
 +  2 (\l\g^{mk^{1}}H^{n}_{1}) F^{2}_{mn} i \ap ^2 &\Vuddef\cr&
 -  2 B^{1}_{mnp} (\l \g^{mnp}  W^2) \ap
 +  36 G^{1}_{k^{2}m} (\l \g^{m}  W^2) i \ap ^2
 }$$
that satisfies
\eqn\QVuot{
QV^{(1,0)}_{1 2} = \ap(k_1+k_2)^2\big(2\ap (\l H^m_1)(\l\g^m W_2)\big)\,.
}
Therefore, a BRST-equivalent Berends-Giele current $M^{(1,0)}_{12}$ can be taken to be
\eqn\BGB{
M^{(1,0)}_{12}={V^{(1,0)}_{1 2}\over\ap(k_1+k_2)^2}.
}

\newsubsec\bgtwosec Berends-Giele current for two massive multiplets

We are now going to explicitly solve the equations \BGsm\ and \bimplies\ for the case
of two massive multiplets
\eqn\QMuut{
M^{(1,1)}_{1 2}={[\Vo1,U^{(2)}_2]_3\over \ap(k_1+k_2)^2}\,,\quad
QM^{(1,1)}_{1 2} = [\Vo1, \Vo2]_2,
}
with $[\Vo1, \Vo2]_2$ given by \dpole\ and $U^{(2)}$ given by \intver.

A long but straightforward calculation using the massive vertex operators \Vdefs\ and \intver\
shows that
\eqnn\triple
$$\eqalignno{
 [\Vo1, U^{(2)}_2]_3 & =
 -  72 (\l H^{k^{2}}_{1}) G^{2}_{k^{1}k^{1}} i \ap ^3 
 +  {177 \over 2}\, (\l H^{m}_{1}) G^{2}_{k^{1}m} i \ap ^2 &\triple\cr&
 -  {45 \over 4}\, (\l \g^{k^{1}k^{2}mn} H_{1}^{p}) B^{2}_{mpn} \ap ^2 
 +  3 (\l \g^{k^{1}k^{2}mn} H_{1}^{p}) B^{2}_{mpn} (k^1 \cdot k^2) \ap ^3 \cr&
 -  36 (\l \g^{k^{1}k^{2}} H_{1}^{m}) G^{2}_{k^{1}m} i \ap ^3 
 -  {17 \over 4}\, (\l \g^{k^{1}mnp} H_{1}^{k^{2}}) B^{2}_{mnp} \ap ^2 \cr&
 +  (\l \g^{k^{1}mnp} H_{1}^{k^{2}}) B^{2}_{mnp} (k^1 \cdot k^2) \ap ^3 
 +  36 (\l \g^{k^{1}m} H_{1}^{k^{2}}) G^{2}_{k^{1}m} i \ap ^3 \cr&
 -  44 (\l \g^{k^{1}m} H_{1}^{n}) B^{2}_{k^{1}nm} \ap ^2 
 +  16 (\l \g^{k^{1}m} H_{1}^{n}) B^{2}_{k^{1}nm} (k^1 \cdot k^2) \ap ^3 \cr&
 -  {33 \over 2}\, (\l \g^{k^{1}m} H_{1}^{n}) G^{2}_{mn} i \ap ^2 
 +  (\l \g^{k^{2}mnp} H_{1}^{k^{2}}) B^{2}_{mnp} \ap ^2 \cr&
 -  30 (\l \g^{k^{2}m} H_{1}^{k^{2}}) G^{2}_{k^{1}m} i \ap ^3 
 +  {11 \over 2}\, (\l \g^{k^{2}m} H_{1}^{n}) B^{2}_{k^{1}nm} \ap ^2 \cr&
 +  2 (\l \g^{k^{2}m} H_{1}^{n}) B^{2}_{k^{1}nm} (k^1 \cdot k^2) \ap ^3 
 +  18 (\l \g^{k^{2}m} H_{1}^{n}) G^{2}_{mn} i \ap ^2 \cr&
 -  6 (\l \g^{k^{2}m} H_{1}^{n}) G^{2}_{mn} (k^1 \cdot k^2) i \ap ^3 
 +  {55 \over 4}\, (\l \g^{mn} H_{1}^{k^{2}}) B^{2}_{k^{1}mn} \ap ^2 \cr&
 -  (\l \g^{mn} H_{1}^{k^{2}}) B^{2}_{k^{1}mn} (k^1 \cdot k^2) \ap ^3 
 +  5 (\l \g^{mn} H_{1}^{p}) B^{2}_{mpn} \ap  \cr&
 +  {9 \over 4}\, (\l \g^{mn} H_{1}^{p}) B^{2}_{mpn} (k^1 \cdot k^2) \ap^2 
 -  (\l \g^{mn} H_{1}^{p}) B^{2}_{mpn} (k^1 \cdot k^2)^2 \ap ^3 \cr&
 +  15 B^{1}_{k^{2}mn} (\l \g^{k^{1}k^{2}mn} H_{2}^{k^{1}}) \ap ^3 
 +  24 B^{1}_{k^{2}mn} (\l \g^{k^{1}n} H_{2}^{m}) \ap ^2 \cr&
 -  12 B^{1}_{k^{2}mn} (\l \g^{k^{1}n} H_{2}^{m}) (k^1 \cdot k^2) \ap ^3 
 -  8 B^{1}_{k^{2}mn} (\l \g^{k^{2}n} H_{2}^{m}) \ap ^2 \cr&
 +  16 B^{1}_{k^{2}mn} (\l \g^{k^{2}n} H_{2}^{m}) (k^1 \cdot k^2) \ap ^3 
 -  30 B^{1}_{k^{2}mn} (\l \g^{mn} H_{2}^{k^{1}}) \ap ^2 \cr&
 +  18 B^{1}_{k^{2}mn} (\l \g^{mn} H_{2}^{k^{1}}) (k^1 \cdot k^2) \ap ^3 
 +  6 B^{1}_{mnp} (\l \g^{k^{1}k^{2}mp} H_{2}^{n}) \ap ^2 \cr&
 -  {5 \over 4}\, B^{1}_{mnp} (\l \g^{k^{1}mnp} H_{2}^{k^{1}}) \ap ^2 
 +  2 B^{1}_{mnp} (\l \g^{k^{1}mnp} H_{2}^{k^{1}}) (k^1 \cdot k^2) \ap ^3 \cr&
 +  4 B^{1}_{mnp} (\l \g^{k^{2}mnp} H_{2}^{k^{1}}) \ap ^2 
 +  B^{1}_{mnp} (\l \g^{k^{2}mnp} H_{2}^{k^{1}}) (k^1 \cdot k^2) \ap ^3 \cr&
 -  {55 \over 4}\, B^{1}_{mnp} (\l \g^{mp} H_{2}^{n}) \ap  
 -  4 B^{1}_{mnp} (\l \g^{mp} H_{2}^{n}) (k^1 \cdot k^2) \ap ^2 \cr&
 +  6 B^{1}_{mnp} (\l \g^{mp} H_{2}^{n}) (k^1 \cdot k^2)^2 \ap ^3 
 +  42 G^{1}_{k^{2}k^{2}} (\l H^{k^{1}}_{2}) i \ap ^3 \cr&
 -  54 G^{1}_{k^{2}m} (\l H^{m}_{2}) i \ap ^2 
 -  6 G^{1}_{k^{2}m} (\l H^{m}_{2}) (k^1 \cdot k^2) i \ap ^3 \cr&
 -  42 G^{1}_{k^{2}m} (\l \g^{k^{2}m} H_{2}^{k^{1}}) i \ap ^3 
 +  {33 \over 2}\, G^{1}_{mn} (\l \g^{k^{1}m} H_{2}^{n}) i \ap ^2 \cr&
 +  18 G^{1}_{mn} (\l \g^{k^{2}m} H_{2}^{n}) i \ap ^2 
 +  6 G^{1}_{mn} (\l \g^{k^{2}m} H_{2}^{n}) (k^1 \cdot k^2) i \ap ^3 \cr&
}$$
indeed satisfies \QMuut. This can be simplified if we define
\eqn\newdefs{
M^{(1,1)}_{1 2}={[\Vo1,U^{(2)}_2]'_3\over \ap(k_1+k_2)^2}\,,\quad
[\Vo1, U^{(2)}_2]'_3 = [\Vo1,U_2^{(2)}]_3 -\ap(k_1+k_2)^2 \Lambda_{12}\,,
}
where $\Lambda_{12}$ is given in \Lambdadef. Now we have
\eqn\QMt{
QM^{(1,1)}_{1 2} = [\Vo1, \Vo2]'_2,
}
with $[\Vo1, \Vo2]'_2$ given by \tptS.

\appendix{E}{OPE brackets in the pure spinor formalism}
\applab\psopeapp

\noindent 
For reviews of the pure spinor formalism, see \refs{\ICTP,\treereview}.
\eqnn\allopes
$$\displaylines{
\p\t^\a(z)\big\{\p\t^\b(w),\Pi^m(w), N^{mn}(w)\big\} \sim{\rm regular},\qquad
[d_\a,\p\t^\b]_2 =  \d^\b_\a,\hfil\allopes\hfilneg\cr
[d_\a, K]_1 =  D_\a K, \quad
[\Pi^m, K]_1 = - 2\ap \p^m K, \quad
[d_\a, \Pi^m]_1 =  (\g^m\p\t)_\a\cr
[d_\a,d_\b]_1 = -  {1\over 2\ap}\g^m_{\a\b}\Pi_m, \quad
[\Pi^m,\Pi^n]_2 = -  2\ap \eta^{mn}, \quad
[d_\a,\t^\b]_1 =  \d^\b_\a\cr
[J,J]_2= - 4,\quad
[J,\l^\a]_1 =  \l^\a,\quad
[N^{mn},\l^\a]_1 = \half (\g^{mn}\l)^\a,\cr
[N^{mn},N^{pq}]_1 = \delta^{p[m} N^{n]q}  - \delta^{q[m} N^{n]p},\quad
[N^{mn},N^{pq}]_2 = -3\delta^{m[q}\delta^{p]n}
}$$
where $K$ is a superfield that does not depend on derivatives of $X^m$ and
$\t^\a$.

\ninepoint
\listrefs

\bye